\documentclass[%
 aip,
 amsmath,amssymb,
 reprint,%
]{revtex4-1}

\usepackage{placeins}

\usepackage{amssymb}
\usepackage{amsmath,textcomp} 
\usepackage{bm}
\usepackage{amstext}
\usepackage{booktabs}
\usepackage{hhline}

\usepackage{graphicx}
\usepackage{dcolumn}
\usepackage{bm}

\usepackage[utf8]{inputenc}
\usepackage[T1]{fontenc}
\usepackage{mathptmx}
\usepackage{etoolbox}
\usepackage{array}
\newcolumntype{L}[1]{>{\raggedright\arraybackslash}p{#1}}

\usepackage[nohyperlinks,nolist]{acronym}
\usepackage{pstricks}
\usepackage{transparent}
\usepackage{comment}
\usepackage{wasysym}
\usepackage{pst-3dplot}
\usepackage{pst-tree}
\usepackage{pst-gr3d}
\usepackage{tikz}
\usetikzlibrary{arrows,calc,positioning,shadows,shapes,shapes.geometric}
\usepackage{pgfplots}
\usepackage{tabularx}
\usepackage{colortbl}
\usepackage{mathtools}
\usepackage{diagbox}
\usepackage{multirow}
\usepackage{eurosym}

\usepackage{siunitx}
\DeclareSIUnit{\px}{px}

\usepackage{import}

\usepackage{caption}
\usepackage{subcaption}

\makeatletter
\def\@email#1#2{%
 \endgroup
 \patchcmd{\titleblock@produce}
  {\frontmatter@RRAPformat}
  {\frontmatter@RRAPformat{\produce@RRAP{*#1\href{mailto:#2}{#2}}}\frontmatter@RRAPformat}
  {}{}
}%
\makeatother
\begin{document}

\preprint{AIP/123-QED}

\title[Comparative Analysis of the Flow in a Realistic Human Airway]{Comparative Analysis of the Flow in a Realistic Human Airway}

\author{Mario Rüttgers}
\affiliation{Data-Driven Fluid Engineering (DDFE) Laboratory, Inha University, Incheon, Republic of Korea}
\author{Julian Vorspohl}
\affiliation{Institute of Aerodynamics and Chair of Fluid Mechanics (AIA), RWTH Aachen University, Aachen, Germany}
\author{Luca Mayolle}
\affiliation{Institute of Aerodynamics and Chair of Fluid Mechanics (AIA), RWTH Aachen University, Aachen, Germany}
\author{Benedikt Johanning-Meiners}
\affiliation{Institute of Aerodynamics and Chair of Fluid Mechanics (AIA), RWTH Aachen University, Aachen, Germany}
\author{Dominik Krug}
\affiliation{Institute of Aerodynamics and Chair of Fluid Mechanics (AIA), RWTH Aachen University, Aachen, Germany}
\author{Michael Klaas}
\affiliation{Institute of Aerodynamics and Chair of Fluid Mechanics (AIA), RWTH Aachen University, Aachen, Germany}
\author{Matthias Meinke}
\affiliation{Institute of Aerodynamics and Chair of Fluid Mechanics (AIA), RWTH Aachen University, Aachen, Germany}
\author{Sangseung Lee}
\affiliation{Data-Driven Fluid Engineering (DDFE) Laboratory, Inha University, Incheon, Republic of Korea}
\author{Wolfgang Schröder}
\affiliation{Institute of Aerodynamics and Chair of Fluid Mechanics (AIA), RWTH Aachen University, Aachen, Germany}
\author{Andreas Lintermann}
\affiliation{Jülich Supercomputing Centre (JSC), Forschungszentrum Jülich GmbH, Jülich, Germany}

\email{m.ruettgers@inha.ac.kr}
\email{a.lintermann@fz-juelich.de}

\date{\today}

\begin{abstract}
Accurate simulations of the flow in the human airway are essential for advancing diagnostic methods. 
Many existing computational studies rely on simplified geometries or turbulence models, limiting their simulation's ability to resolve flow features such shear-layer instabilities or secondary vortices. 
In this study, direct numerical simulations were performed for inspiratory flow through a detailed airway model which covers the nasal mask region to the 6th bronchial bifurcation. 
Simulations were conducted at two physiologically relevant \textsc{Reynolds} numbers with respect to the pharyngeal diameter, i.e., at
$Re_p=400$ (resting) and $Re_p=1200$ (elevated breathing).
A lattice-Boltzmann method was employed to directly simulate the flow, i.e., no turbulence model was used. 
The flow field was examined across four anatomical regions: 1) the nasal cavity, 2) the naso- and oropharynx, 3) the laryngopharynx and larynx, and 4) the trachea and carinal bifurcation.
The total pressure loss increased from $9.76~Pa$ at $Re_p=400$ to $41.93~Pa$ at $Re_p=1200$. 
The nasal cavity accounted for the majority of this loss for both \textsc{Reynolds} numbers, 
though its relative contribution decreased from $81.3\%$ at $Re_p=400$ to $73.4\%$ at $Re_p=1200$.
At $Re_p=1200$, secondary vortices in the nasopharyngeal bend and turbulent shear-layers in the glottis jet enhanced the local pressure losses. 
In contrast, the carinal bifurcation mitigated upstream unsteadiness and stabilized the flow.
A key outcome is the spatial correlation between the pressure loss and the onset of flow instabilities across the four regions. This yields a novel perspective on how the flow resistance and vortex dynamics vary with geometric changes and flow rate. 
\end{abstract}

\maketitle

\begin{acronym}
\acro{3D-PTV}[3D-PTV]{3D particle-tracking velocimetry}
\acro{ANN}[ANN]{Artificial Neural Network}
\acro{BGK}[BGK]{Bhatnagar–Gross–Krook}
\acro{CFD}[CFD]{computational fluid dynamics}
\acro{CAD}[CAD]{Computer Aided Design}
\acro{CPU}[CPU]{Central Processing Unit}
\acro{CNN}[CNN]{convolutional neural network}
\acro{CNPAS}[CNPAS]{congenital nasal pyriform aperture stenosis}
\acro{CT}[CT]{computed tomography}
\acro{DICOM}{digital imaging and communications in medicine}
\acro{CNN}[CNN]{convolutional neural network}
\acro{DNS}[DNS]{direct numerical simulation}
\acro{GPU}[GPU]{graphics processing unit}
\acro{GIS}[GIS]{geographic information system}
\acro{GNN}[GNN]{Graph Neural Network}
\acro{GUI}[GUI]{graphical user interface}
\acro{HPC}[HPC]{High-Performance Computing}
\acro{IGA}[IGA]{isogeometric analysis}
\acro{JURECA-DC}[JURECA-DC]{Jülich Research on Exascale Cluster Architectures}
\acro{JSC}[JSC]{Jülich Supercomputing Centre}
\acro{LB}[LB]{lattice-Boltzmann}
\acro{LBM}[LBM]{Lattice-Boltzmann Method}
\acro{LES}[LES]{large-eddy simulation}
\acro{LiDAR}[LiDAR]{Light Detection and Ranging}
\acro{m-AIA}[m-AIA]{multiphysics-Aerodynamisches Institut Aachen}
\acro{MAPE}[MAPE]{mean absolute percentage error}
\acro{MaxAPE}[MaxAPE]{maximum absolute percentage error}
\acro{MARME}[MARME]{miniscrew-assisted rapid maxillary expansion}
\acro{ML}[ML]{Machine Learning}
\acro{NetCDF}[NetCDF]{Network Common Data Form}
\acro{NURBS}[NURBS]{non-uniform rational B-splines}
\acro{PPDF}[PPDF]{particle probability distribution function}
\acro{PIV}[PIV]{particle image velocimetry}
\acro{RANS}[RANS]{Reynolds-averaged Navier Stokes}
\acro{ROI}[ROI]{region of interest}
\acro{STL}[STL]{Stereolithography}
\acro{StB}[STB]{shake-the-box}
\acro{SGMS-GNN}[SGMS-GNN]{Subgraph Multi-Scale Graph Neural Network}
\acro{ZFS}[ZFS]{Zonal Flow Solver}
\end{acronym}

\section{Introduction}
\label{sec:intro}
Understanding the flow physics of the airflow in the human airway is essential for applications ranging from inhalation therapy~\cite{Dutta2020,Dastoorian2022} and surgical planning~\cite{Ruettgers2024,Waldmann2022} to modeling mechanical ventilators~\cite{Carson2024}. 
The complexity of the airway which is characterized by intricate geometries with sharp bends and rapidly varying cross sections gives rise to various flow phenomena including laminar-to-turbulent transition, vortex formation, jet-like flow structures, recirculation zones etc. 
Accurately resolving such features is critical for the analysis of, e.g., transport and deposition of aerosols~\cite{Lintermann2017}, predicting airway resistance to quantify respiration inhibition, or ultimately, for physics-based surgery decision making~\cite{Ruettgers2025a,Ruettgers2025}. 

The formation of flow structures is determined by the geometry and flow parameters. 
Despite the importance of the geometry, most studies focus on only a fraction of the airway such as the nasal cavity~\cite{Lintermann2013,Niegodajew2025}, glottis~\cite{Emmerling2024,Yang2020}, or trachea~\cite{Xu2020,Morita2022}. 
Furthermore, in many studies, simplified airway geometries such as straight‐walled sections, idealized \ac{CAD} segments, or even two-dimensional approximations are considered to facilitate meshing and reduce computational cost~\cite{Ruettgers2021,Vivek2024}. 
Comparative analyses reveal that these models often fail to capture essential flow features, including jets, recirculation zones, and regional deposition patterns~\cite{Johari2013,DeBacker2008,Aljawad2021,Nasrul2013}.

Numerical investigations frequently rely on solving the \ac{RANS} equations including turbulence models which are developed for different flow categories. 
Such computations are efficient~\cite{Sommerfeld2021,Saksono2011},
however, they provide only low resolution solutions of a temporally averaged flow field.
In respiratory flow simulations, \ac{RANS} simulations underperform in computing pressure drop~\cite{Schillaci2022} and in representing turbulent fluctuations and oscillatory wall-shear stress, 
which are important to accurately predict aerosol transport and pathology-related flows~\cite{Chen2013}.
Even though some scale-resolving, unsteady \ac{RANS} simulations are able to reproduce the pressure drop in airways,
they show weaknesses in capturing wall-shear stress fluctuations~\cite{Shao2021}.
Thus, despite their efficiency, \ac{RANS} methods contain modeling errors that are critical to study patient-specific airway dynamics and aerosol behavior.

Alternatively, \ac{LES} and \ac{DNS} offer a more detailed analysis of the underlying flow physics.
\ac{LES} resolve the large-scale turbulent eddies directly and the small subgrid-scale motions are modeled.
Hence, unsteady flow structures and spatial heterogeneities can be more accurately captured, however, at significantly higher computational cost compared to \ac{RANS} simulations. 
In respiratory flows, \ac{LES} have been used to understand complex flow features such as recirculation zones, shear-layer instabilities, and secondary vortices that are typically unsufficiently resolved or misrepresented in \ac{RANS} simulations~\cite{Zuber2012,Li2017,Calmet2015}.
Unlike \ac{LES} methods, \ac{DNS} resolve the entire spectrum of spatial and temporal flow scales without any modeling such that the "purest" flow field at even higher computational cost will be determined.

Comparing numerical simulations of respiratory flows with experimental approaches such as \ac{PIV} is crucial to validate numerical predictions~\cite{Xu2020}. 
For example, a carefully checked numerical simulation software can be a powerful tool to characterize the deposition of inhaled particles in realistic geometries of the airway~\cite{Farkas2020,Lizal2020,Sadafi2024}.
In other words, validation even for \ac{LES} and \ac{DNS} is necessary to ensure physical accuracy in patient-specific geometries. 
Thus, extensive high-fidelity simulations must be validated against in vitro or in vivo measurement data to confirm that complex flow structures determined in intricate geometries agree with experimental findings. 

Additionally, airflow is frequently studied at a single flow rate, 
neglecting the variation introduced by different breathing intensities or activity levels~\cite{Hoerschler2010,Lintermann2012}. 
Inhalation during exercise produces qualitatively and quantitatively significantly different flow dynamics from breathing at rest. These differences can strongly influence aerosol transport, pressure distributions, and shear stress distributions. 
For example, the peak velocities, wall pressures, and wall-shear stresses can be up to $9$ times higher when running instead of walking~\cite{Tsega2022}. 
Similarly, variations in the inhalation duration and respiratory rate cause substantial differences in secondary and lateral flow patterns even when peak \textsc{Reynolds} numbers are held constant~\cite{Gaddam2021}.
Note that the \textsc{Reynolds} number is defined by the mean velocity, a relevant length scale, and the kinematic viscosity of air, i.e., it is proportional to the mean flow rate.
The findings indicate that simulations at a single flow condition may miss critical variations in flow structures occurring in real-world respiratory scenarios, such as asymmetries or transitional turbulence, that are important to accurately predict drug delivery, airway collapse risk, patient-specific treatment planning etc. 
Therefore, a comprehensive analysis should include multiple breathing conditions, e.g., rest and exertion, to cover the functional range of the airway physiology and to validate \ac{CFD} simulations appropriately.

\begin{figure}
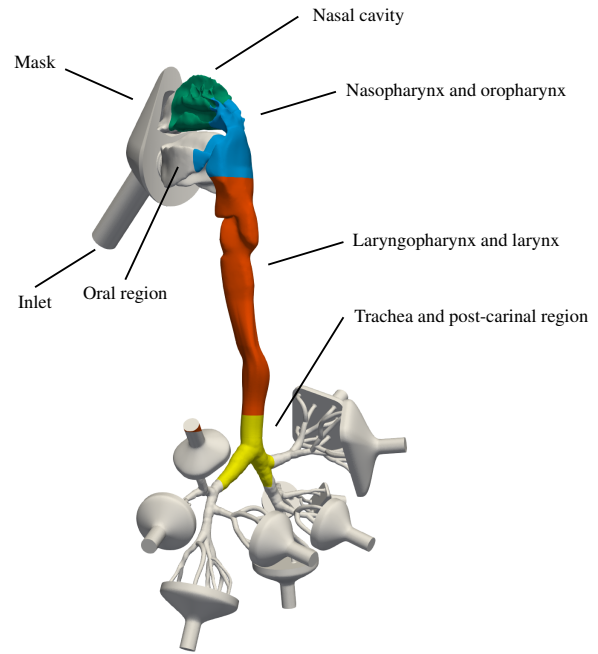

\include{figures/model}
\caption{3D model of the human airway~\cite{Farkas2020,Lizal2020}.}
\label{fig:model}
\end{figure}

The current study addresses the previously mentioned limitations. 
\ac{DNS} of the flow at inhalation for a realistic, adult human airway geometry from a mask covering the nostrils to the 6th bronchial bifurcation are performed.
The geometry is shown in Fig.~\ref{fig:model}.
The connection between the mask and the oral region is closed and only nasal breathing is considered.
Simulations are conducted at the \textsc{Reynolds} numbers $Re = 400$ and $Re = 1200$ to capture the flow structures of peak inhalation conditions under resting and moderately increased breathing. 

The \ac{DNS} is based on a high-fidelity \ac{LB} method using $435$ million cells such that a fully resolved representation of the unsteady flow phenomena is achieved. 
The \ac{LB} method is part of the open-source simulation framework \ac{m-AIA}~\cite{m-AIA}. 
It has been used for numerous numerical investigations of nasal cavity flows such as thermal analyses of patients' breathing capabilities~\cite{Lintermann2013,Waldmann2020,Lintermann2019},
machine learning-based automatization of pre-processing and executing \ac{CFD} simulations~\cite{Waldmann2021,Ruettgers2022,Liu2024},
and automated virtual surgery planning for treating respiratory diseases~\cite{Ruettgers2024,Ruettgers2025a,Ruettgers2025}.

The \ac{DNS} results are validated against high-resolution experimental data measured in the same geometry.
The experiments were performed using a surface-coated biological model and a colorimetric {3-(4,5dimethylthiazol-2-yl)-2,5-diphenyltetrazolium} bromide (MTT) assay to determine the aerosol deposition in the airways. 
First, high-speed mono-\ac{PIV} measurements were conducted to determine the relationship between the flow field and the aerosol deposition~\cite{JohanningMeiners.2023}. 
Second, to investigate the influence of oral and nasal inhalation on the lower airways at peak in- and exhalation, 
\ac{3D-PTV} measurements using the \ac{StB} algorithm were performed~\cite{JohanningMeiners.2024}.

While the experimental investigations are limited to accessible regions, 
\ac{DNS} yield data for the entire domain revealing flow features not observable in the experiments.
The analysis in this study focuses on the development and evolution of flow structures in the following four key anatomical regions
which are highlighted by different colors in Fig.~\ref{fig:model}:
\begin{itemize}
    \setlength{\itemsep}{-0.2em}
    \item The nasal cavity (green), which is characterized by narrow valve regions and turbinate induced flow acceleration with complex, multi-scale recirculation.
    \item The naso- and oropharynx (blue), where curvature induced vortex pairs develop due to the nasal–pharyngeal transition.
    \item The laryngopharynx and larynx (red), featuring shear-layer instabilities and jet-like behavior frequently referred to as “glottis jet".
    \item The trachea and post-carinal regions (yellow), where downstream stabilization and interaction of secondary vortices occur.
\end{itemize}

To the best of the authors' knowledge, this is the first validated \ac{DNS} study of the complex airflow in a realistic human airway model covering the full inhalation path down to the 6th bifurcation at two \textsc{Reynolds} numbers. 
The results highlight how vortical structures emerge, interact, and evolve across anatomical boundaries.
Transition mechanisms and secondary flow features are discussed that are absent or not resolved in previous studies. 
A further novel aspect of this study is the systematic, region-by-region correlation analysis of the pressure loss which is complemented by considering flow instabilities and transitional behavior. 
This comprehensive analysis offers a new perspective on how geometry and flow rate influence respiratory airflow and is a further step for improved physiological modeling and diagnostic approaches.

This manuscript is structured as follows.
Section~\ref{sec:methods} presents
the numerical and experimental techniques used in this study.
In Sec.~\ref{sec:results}, the results of a mesh refinement study including a comparison to experimental data are shown 
and the flow field in the four key anatomical regions is analyzed.
Then, the results are summarized, discussed, and an outlook is given in Sec.~\ref{sec:disc}.

\section{Methods}
\label{sec:methods}
The numerical methods (Sec.~\ref{sec:num}) including the boundary conditions and experimental techniques (Sec.~\ref{sec:exp}) are presented.

\subsection{Numerical methods}
\label{sec:num}

The numerical simulations are performed using the \ac{LB} method implemented within the \ac{m-AIA} framework. 
The \ac{LB} approach is particularly advantageous for handling highly complex and detailed geometries~\cite{Geller2006}.
The \ac{LB} module solves the discretized form of the Boltzmann
equation with the \ac{BGK} approximation of the right-hand side collision process~\cite{heTheoryLatticeBoltzmann1997}.
That is
\begin{equation}
	f_i(\boldsymbol{x} + \boldsymbol{\xi_i} \delta t, t + \delta t) - f_i(\boldsymbol{x}, t) = - \omega(f_i(\boldsymbol{x}, t)  -  f^{eq}_i(\boldsymbol{x}, t) ),
	\label{eq:BGK-equation}
\end{equation}
is solved for the \acp{PPDF} $f_i$ at neighboring fluid cells at locations $\boldsymbol{x} + \boldsymbol{\xi_i} \delta t$. 
The \acp{PPDF} are functions of the location vector $\boldsymbol{x} = (x_1, x_2, x_3)^T$,
the discrete molecular velocity vector $\boldsymbol{\xi_i}=(\xi_1,\xi_2,\xi_3)^T$, 
and the time $t$ and the time increment $\delta t$.
The collision frequency is expressed by $\omega$.
The discretization is based on the \verb*|D3Q27| model~\cite{qianLatticeBGKModels1992}, with $i \in \{1,\ldots, Q=27\}$ directions in 3D. 
The discrete Boltzmann-Maxwell distribution function reads
\begin{equation}
	{f}^{eq}_i = wc_i  \rho \left(  1 + \frac{\boldsymbol{\xi_i} \cdot \boldsymbol{u}}{c_s^2} +  \frac{1}{2} \left(\frac{\boldsymbol{\xi_i} \cdot \boldsymbol{u}}{c_s^2}\right)^2  -  \frac{\boldsymbol{u} \cdot \boldsymbol{u}}{2c_s^2} \right),
	\label{eq:boltzmann_eq_dist_discrete}
\end{equation}
with the isothermal speed of sound $c_s=1/\sqrt{3}$, fluid density $\rho$, fluid velocity vector $\boldsymbol{u}=(u,v,w)^T$ with the velocity components $u$, $v$, and $w$ in the $x$-, $y$- and $z$-directions, and weight coefficients $wc_i$~\cite{qianLatticeBGKModels1992}.

The macroscopic variables $\rho$ and $\boldsymbol{u}$ can be computed by
\begin{align}
	\rho &= \sum_{i=1}^{Q} f_i , \label{eq:macro_rho} \\
	\rho \boldsymbol{u} &= \sum_{i=1}^{Q} \boldsymbol{\xi}_i \cdot f_i.
\end{align}
The static pressure is obtained by the equation of state $p_{stat}=c_s^2\rho$.

%
\begin{figure}
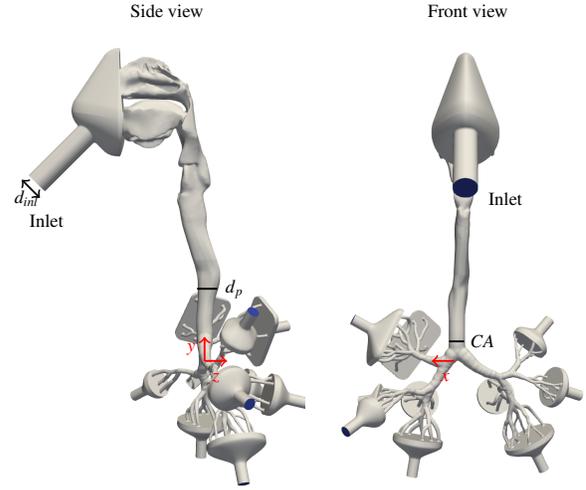

\include{figures/domain}
\caption{Computational domain.}
\label{fig:domain}
\end{figure}
The computational domain and the coordinate system are illustrated in Fig.~\ref{fig:domain}.
The nostrils and mouth are covered by a mask that is extended by a pipe with a diameter of $d_{inl}=\SI{20}{\milli\meter}$.
The pipe has an inclination angle of $\alpha=\SI{45}{\degree}$.
At ten locations, the 5th and 6th bifurcations of the lung are merged into a single pipe outlet with diameter $d_{outl}=\SI{8}{\milli\meter}$, yielding in total ten outlets.
The \textsc{Reynolds} number $Re_p=(U_{p} \cdot d_p)/\nu$ is based on the hydraulic diameter of the pharynx region $d_p=\SI{16.3}{\milli\meter}$ at $y=2.92\cdot d_p$,
the kinematic viscosity $\nu$,
and the spatially averaged velocity at the pharynx $U_{p}$.
To match the measurements of Johanning-Meiners et al.~\cite{JohanningMeiners.2023,JohanningMeiners.2024}, results for $Re_p=400$ and $Re_p=1200$ are investigated.
The inflow velocity $U_{inl}$ is determined by the \textsc{Reynolds} number at the inlet $Re_{inl}=(d_p/d_{inl}) \cdot Re_p$.
The values of the quantities $\nu$, $U_p$, and $U_{inl}$ for air, which are used to compute the 
dimensional 
total pressure loss in Sec.~\ref{sec:res_3}, are given in Tab.~\ref{tab:flow_param}.

\begin{table*}[]
\centering
\renewcommand{\arraystretch}{1.2}
\begin{tabular}{c|c|cc|cc}
\hline
\multirow{2}{*}{Fluid} & \multirow{2}{*}{$\nu$ ($\si{\square\meter\per\second}$)} & \multicolumn{2}{c|}{$Re_p = 400 ~(I)$} & \multicolumn{2}{c}{$Re_p = 1200 ~(II)$} \\
 & & $U_p^I$ ($\si{\meter\per\second}$) & $U_{inl}^I$ ($\si{\meter\per\second}$) & $U_p^{II}$ ($\si{\meter\per\second}$) & $U_{inl}^{II}$ ($\si{\meter\per\second}$) \\
\hline
Air & $1.63 \cdot 10^{-5}$ & 0.49 & 0.4 & 1.47 & 1.2 \\
Water+Glycerin & $5.44 \cdot 10^{-6}$ & 0.13 & 0.09 & 0.39 & 0.27 \\
\hline
\end{tabular}
\vspace{2mm}
\caption{Quantities $\nu$, $U_p$, and $U_{inl}$ for different fluids and \textsc{Reynolds} numbers.}
\label{tab:flow_param}
\end{table*}

%
%

At the inlet, the velocity profile of a fully developed pipe flow is prescribed
and the density of the fluid is linearly extrapolated from the neighboring inner cells.
At the outlets, a constant pressure $p_{outl}$ is set 
and the velocity is extrapolated from the inner cells.
To satisfy the no-slip condition on the walls,
an interpolated bounce-back scheme is used~\cite{bouzidiMomentumTransferBoltzmannlattice2001}.

The discrete formulation is based on unstructured, hierarchical Cartesian meshes with uniform refinement which are generated using the massively parallel mesh generator integrated in the \ac{m-AIA} framework~\cite{Lintermann2014}. The mesh has an octree data structure, which is created by recursively subdividing an initial bounding cube that encloses the \ac{ROI}~\cite{Hartmann2008}, i.e., the airway. On each refinement level parent cubes are divided into eight smaller sub-cubes, forming a hierarchical structure through well-defined parent-child relationships. Cells outside the \ac{ROI} are discarded to reduce computational overhead and to optimize the mesh.

The mesh is refined up to a pre-defined level, which is used to perform domain decompositioning with a Hilbert space-filling curve~\cite{Sagan1994}. Cells on lower refinement levels are ordered using a Z-curve (Morton order)~\cite{Morton1966}. Concerning parallel processing, the Hilbert curve preserves spatial locality, ensuring that neighboring cells in the domain are typically assigned to the same or adjacent computational processes. This, together with the Z-curve on a sub-domain level, promotes efficient data locality both within and across compute nodes, minimizing communication overhead. Since the domain partitioning is balanced from the outset and remains static, load imbalance does not occur under uniform refinement. 
The final mesh is stored using the parallel \ac{NetCDF} and written via concurrent I/O routines~\cite{Li2003}, enabling scalable and efficient data handling for large simulations.

\subsection{Experimental methods}
\label{sec:exp}
\begin{figure}
\centering
\def\svgwidth{8cm}
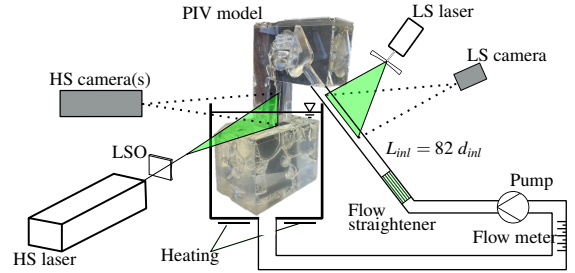
\caption{Sketch of the experimental setup for \ac{PIV} and \ac{3D-PTV} measurements.}
\label{fig:setup_exp}
\end{figure}

The experiments are conducted using a silicone model of the human respiratory tract embedded in a closed-circuit facility mimicking steady inhalation.
The quantities $\nu$, $U_p$, and $U_{inl}$ for the water and glycerin fluid are summarized in Tab.~\ref{tab:flow_param}.
The experimental setup is sketched in Fig.~\ref{fig:setup_exp}.
The silicone's model refractive index of $n= 1.406$ is matched by a mixture of water and glycerine, which is held constant during all the experiments at a temperature of \SI{30}{\celsius}. A fully developed inlet velocity profile is generated by using an inlet pipe with $L_{inl} = 82\cdot d_{inl}$ including a flow straightener downstream of the pump.

Two independent \ac{PIV} systems are used in the experiments. 
For the first system, the velocity profile upstream of the mask and thus the \textsc{Reynolds} number in the trachea is measured using a Continuum Minilite low-speed laser combined with a pco.edge sCMOS camera. Two-dimensional velocity fields are captured using a Darwin-Duo \SI{40}{\milli\joule} laser and a Photron FASTCAM NOVA S12, achieving a spatial resolution of $\SI{17}{\px\per\milli\meter}$. The raw images are processed using an in-house cross-correlation algorithm and the method is further referred to as \ac{PIV}.

For the second system, two additional Photron FASTCAM Mini WX100 cameras are positioned at an angle of \SI{25}{\degree} resulting in a spatial resolution of $\SI{19.5}{\px\per\milli\meter}$ to perform \ac{3D-PTV} measurements. A Darwin-Duo \SI{100}{\milli\joule} laser with volumetric light sheet optics illuminates the trachea measurement volume. The images obtained from the three cameras are processed in DaVis~10 by LaVision using the \ac{StB} algorithm~\cite{Schanz2016}.

\section{Results}
\label{sec:results}
In Sec.~\ref{sec:res_1}, results of a mesh refinement study are given, which include a validation with the experimental measurements.
Section~\ref{sec:res_2} provides a detailed analyses of flow structures in the four key anatomical regions, see Fig.~\ref{fig:model}, including the effect of the different \textsc{Reynolds} numbers on the flow fields.
In Sec.~\ref{sec:res_3}, the influence of the different flow phenomena on the pressure distribution is analyzed.

The simulations were conducted on the \ac{CPU} partition on the JURECA-DC supercomputer~\cite{JURECA} of the \ac{JSC}, Forschungszentrum Jülich, Germany.
Each node contains two AMD EPYC 7742 processors with 64 cores each, clocked at 2.25 GHz,
and 512 GB DDR4 memory.

\begin{figure*}[]
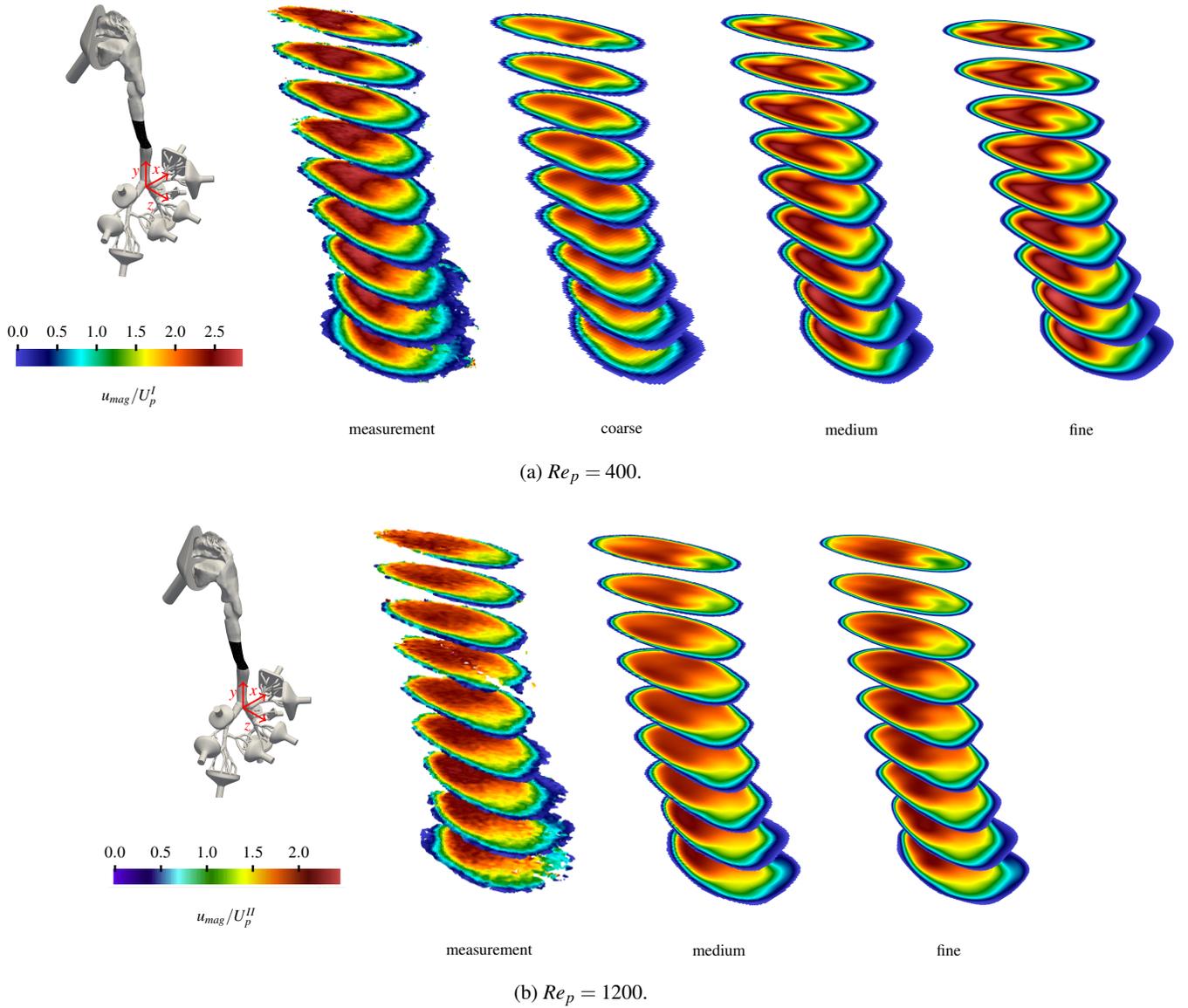

\captionsetup[subfigure]{aboveskip=8pt,belowskip=8pt}
\begin{subfigure}{\linewidth}
\centering
\include{figures/slices_re_400}
\captionsetup{format=hang}
\caption{$Re_p=400$.}
\label{fig:slices_re_400}
\end{subfigure}
\begin{subfigure}{\linewidth}
\centering
\include{figures/slices_re_1200}
\captionsetup{format=hang}
\caption{$Re_p=1200$.}
\label{fig:slices_re_1200}
\end{subfigure}
\caption{Equidistant cross sections of the normalized velocity magnitude $u_{mag}/U_p$ at two \textsc{Reynolds} numbers in the $x$-$z$-plane ranging from $y/d_p=4$ to $y/d_p=6.4$, which is highlighted by the black area in the 3D model on the left in Fig.~\ref{fig:slices_re_1200}. Results for various mesh resolutions are compared to excerpts of the \ac{3D-PTV} measurement series described in~\cite{JohanningMeiners.2024}.}
\label{fig:slices}
\end{figure*}

\subsection{Mesh refinement and validation}
\label{sec:res_1}

Figure~\ref{fig:slices} shows the results of a mesh refinement study at $Re_p=400$ and $Re_p=1200$.
Simulations with a coarse (subscript $c$), medium (subscript $m$), and fine (subscript $f$) mesh at $\Delta x_c=d_{inl}/50$ with $7 \cdot 10^6$ cells, $\Delta x_m=d_{inl}/100$ with $55 \cdot 10^6$ cells,
and $\Delta x_f=d_{inl}/200$ and $433 \cdot 10^6$ cells have been conducted.
The numerical results for the coarse, medium, and fine meshes are compared to data of the \ac{3D-PTV} measurement series described in~\cite{JohanningMeiners.2024}.
Holes in the measurement plots and areas not covering the entire measurement volume are largely due to near-wall reflections and the settings of the \ac{3D-PTV} evaluation.

%
%

%
\begin{figure*}[]
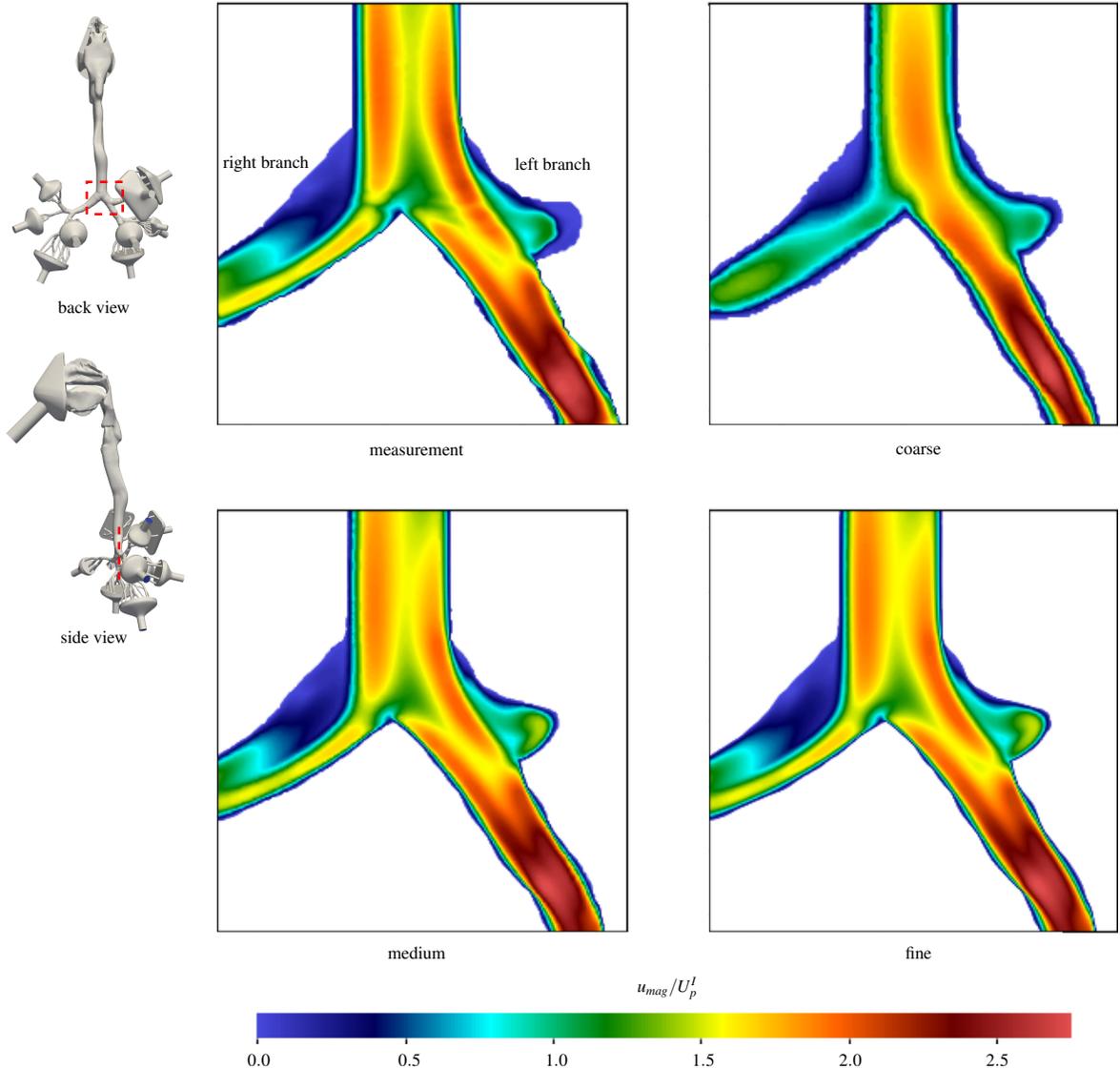

    \centering
    \include{figures/comp_400}
    \caption{Contours of the normalized velocity magnitude $u_{mag}/U_p^I$ at $Re_p=400$ in the bifurcation (dashed red square and dashed red line for $z=0$ in the sketch). Results for the coarse, medium, and fine meshes are compared to \ac{PIV} measurements~\cite{JohanningMeiners.2023}.}
    \label{fig:comparison_400}
\end{figure*}

In Fig.~\ref{fig:slices_re_400}, the results for $Re_p=400$ in the form of nine equidistant cross sections in the $x$-$z$-plane ranging from $y/d_p=4$ to $y/d_p=6.4$ are depicted. 
This region is highlighted by the black area in the 3D model shown on the left side.
The cross sections are colored by the absolute velocity $u_{mag}=\sqrt{u^2+v^2+w^2}$.

All cross sections show a high-speed jet in the center.
The coarse simulation computes a lower velocity magnitude in the center of the jet compared to the measurement. 
In contrast, the numerical velocity magnitude matches well with the experimental value in the medium and fine simulations.
The jet develops a dented shape from the center in the positive $z$-direction.
In the coarse simulation, this dented shape of the jet is only indicated in the upper two cross sections.
In the remaining cross sections, this structure is no longer observed.
Furthermore, while the low velocity part of the jet extends into the center of the cross sections in the measurement,
it is only present near the wall in the coarse simulation.
In other words, the coarse resolution is not sufficient to capture even such large-scale flow features.
However, the simulations based on the medium and fine meshes do reproduce the dented shape and the low velocity part which extends into the center in all cross sections.

In Fig.~\ref{fig:slices_re_1200}, the numerical results for $Re_p=1200$ for the medium and fine meshes are compared with the \ac{3D-PTV} measurement results.
Unlike the $Re_p=400$ case, where the slowly flowing part extended into the center for all cross sections,
for $Re_p=1200$, this is only observed for the first three cross sections in the measurements and the simulations.
It can be stated that the agreement in the development of the flow structures between the measurements and the medium and fine simulations is good.

\begin{figure*}[]
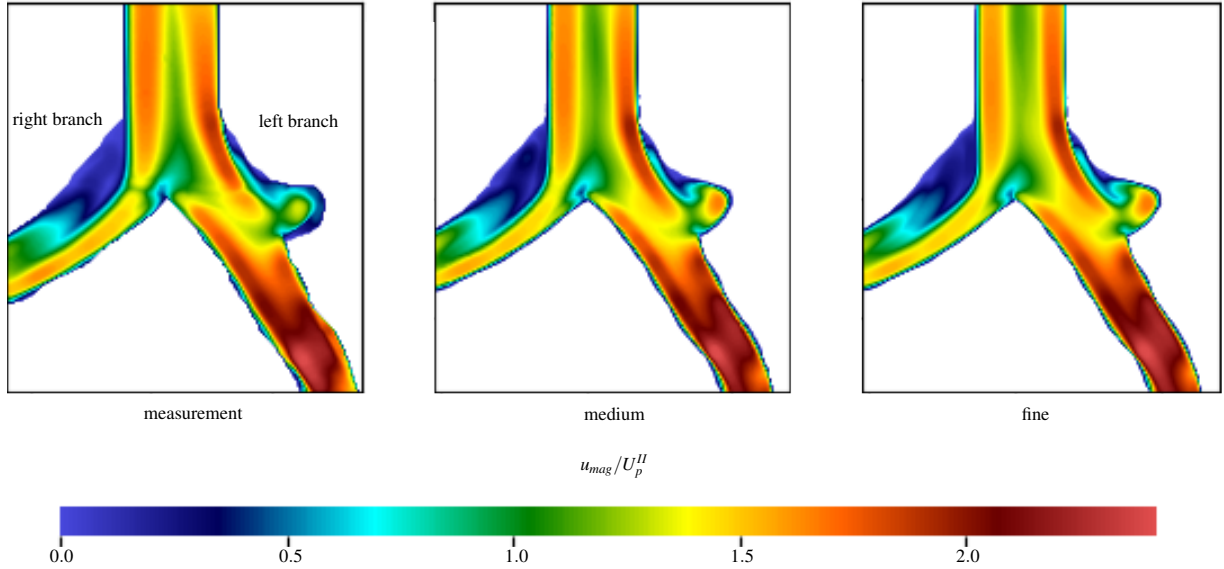

    \centering
    \include{figures/comp_1200}
    \caption{Contours of the normalized velocity magnitude $u_{mag}/U_p^{II}$ at $Re_p=1200$ in the bifurcation (dashed red square and dashed red line for $z=0$ in the sketch in Fig.~\ref{fig:comparison_400}). Results for the medium and fine meshes are compared to \ac{PIV} measurements~\cite{JohanningMeiners.2023}.}
    \label{fig:comparison_1200}
\end{figure*}
\begin{figure*}[]
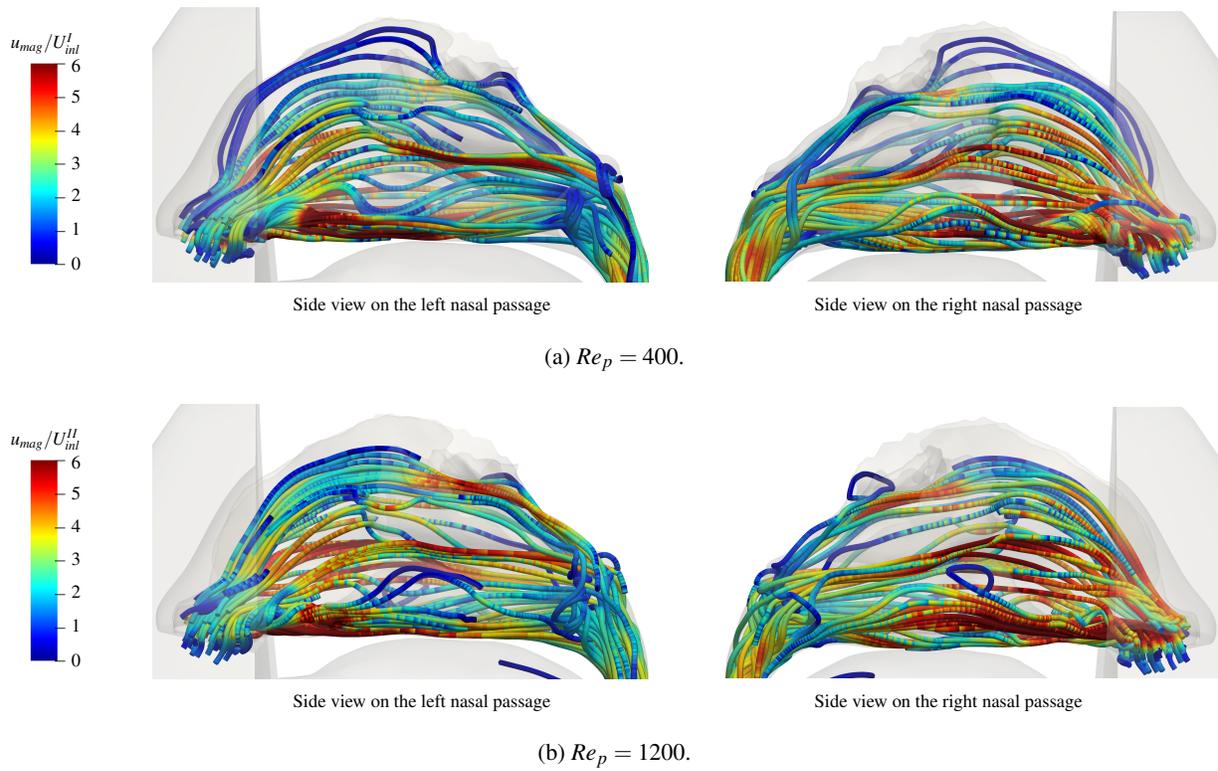

\captionsetup[subfigure]{aboveskip=8pt,belowskip=8pt}
\begin{subfigure}{\linewidth}
\centering
\include{figures/nc_400}
\caption{$Re_p=400$.}
\label{fig:nc_streamlines_400}
\end{subfigure}
\begin{subfigure}{\linewidth}
\centering
\include{figures/nc_1200}
\caption{$Re_p=1200$.}
\label{fig:nc_streamlines_1200}
\end{subfigure}
\caption{Streamlines of the time averaged velocity field in the nasal cavity for $Re_p=400$ and $Re_p=1200$ colored by the normalized velocity magnitude $u_{mag}/U_{inl}$.}
\label{fig:nc_streamlines}
\end{figure*}

Next, the flow field at the bifurcation is assessed by comparing \ac{PIV} measurements from~\cite{JohanningMeiners.2023} to the \ac{LB} simulations in a cross section at $z=0$ and for $x,y\in[-30mm,+30mm]$. 
Figure~\ref{fig:comparison_400} shows the comparison for $Re_p=400$ again for the coarse, medium, and fine resolutions.
Upstream of the bifurcation, the experiment clearly shows two maxima in the velocity profile. 
They are not visible in the coarse simulation since the numerical diffusion in the coarse simulation dampens the development of such a flow structure and its existence in the streamwise direction. 
In the left branch of the bifurcation, the experimental data show a clear flow separation. 
This recirculation zone is covered by a thin wall jet. 
While the flow separation from the right wall is so massive that it is still visible in the coarse simulation, the wall jet following the impingement and the size of the recirculation zone are not well captured by the coarse mesh. 
In contrast, the medium and fine simulations reproduce these flow structures. 
In the left branch, the coarse simulation predicts a centered jet,
while the experiments as well as the medium and fine simulations show two wall jets that merge into one jet downstream.
Hence, the higher resolutions are necessary to determine the essential flow structures.

For the $Re_p=1200$ problem, the results are shown in Fig.~\ref{fig:comparison_1200}. 
Since the coarse grid findings for $Re_p=400$ were not reliable,
only the medium and fine grid results are shown.
The overall flow structure is comparable to the case of $Re_p=400$. 
For the right branch of the bifurcation, the wall jet is noticeably wider at $Re_p=1200$, which is due to the somewhat smaller and thinner separation bubble.
The comparison with the experimental data shows that the medium and fine resolutions are able to predict the flow structures upstream and downstream of the bifurcation with good accuracy.

\subsection{Flow structures in the four key anatomical regions}
\label{sec:res_2}

In the following, only results obtained using the fine mesh are discussed.
The first key anatomical region is the nasal cavity.
Since the time averaged and instantaneous streamlines at $Re_p=400$ and $Re_p=1200$ are quite similar, 
only the time averaged streamlines are analyzed. 
They are shown in Fig.~\ref{fig:nc_streamlines} and are colored by the velocity magnitude, which is normalized by the inflow velocity.
The flow at $Re_p=1200$ exhibits generally larger normalized velocities compared to the $Re_p=400$ case. 
This is especially observed in the channel below the lower turbinate.
Other than that, the spatial flow structures are visually only slightly different in the two cases. 
The flow spreads across all turbinates.
\begin{figure*}
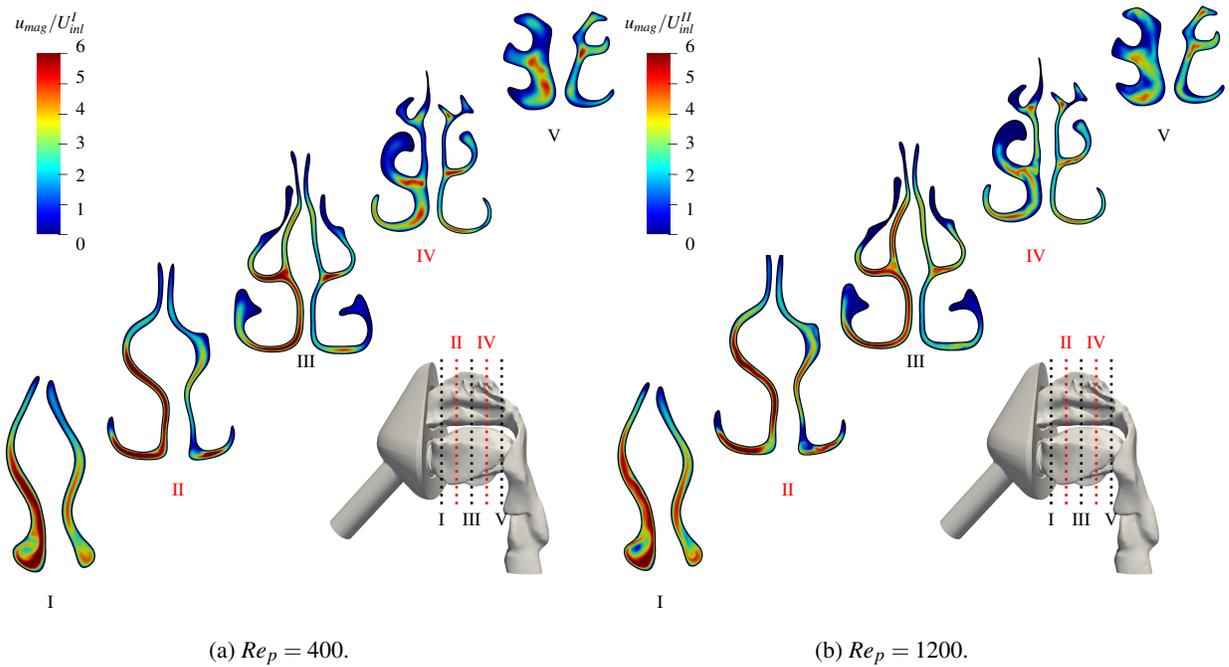

\captionsetup[subfigure]{aboveskip=8pt,belowskip=8pt}
\begin{subfigure}{0.45\linewidth}
\centering
\include{figures/nc_400_cross_sec}
\caption{$Re_p=400$.}
\label{fig:nc_cross_sec_400}
\end{subfigure}
\begin{subfigure}{0.45\linewidth}
\centering
\include{figures/nc_1200_cross_sec}
\caption{$Re_p=1200$.}
\label{fig:nc_cross_sec_1200}
\end{subfigure}
\caption{Time averaged normalized velocity magnitude $u_{mag}/U_{inl}$ at $Re_p=400$ and $Re_p=1200$ in several cross sections I~-~V.}
\label{fig:nc_cross_sec}
\end{figure*}

Figure~\ref{fig:nc_cross_sec} shows $u_{mag}/U_{inl}$ at $Re_p=400$ and $Re_p=1200$ in the cross sections I~-~V, which are defined by the dotted lines in the sketches of the nasal cavity.
At both \textsc{Reynolds} numbers, a notable asymmetry is observed between the left and right nasal passages.
In the cross sections I~-~III, the velocity in the right nasal passage is higher than in the left passage.
In the cross sections IV and V, the right passage is significantly wider than the left,
which leads to a lower velocity and to a more balanced velocity distribution in both nasal passages.
 
The nasal cavity in this study is not pathological, for instance, a deviated septum is not present. 
Therefore, it is assumed that the disparity in the cross sections IV and V are likely to be caused by the nasal cycle. 
The nasal cycle is a natural physiological rhythm during which alternating congestion and decongestion of the turbinates occur in each passage. This leads to a
deceleration and acceleration of the flow through the decongested and the congested sides~\cite{Wei2024}.

The largest difference between the $Re_p=400$ and $Re_p=1200$ cases in Fig.~\ref{fig:nc_cross_sec} is observed in the cross section III.
For the higher \textsc{Reynolds} number, increased velocities not only occur near the inferior turbinate,
but also in the upper channels of the nasal passage.
This implies that increased airflow at higher $Re$ numbers enhances the transport of air into the uppermost regions of the nasal cavity, i.e., 
into the olfactory region.
This region contains specialized olfactory epithelium that houses neural receptors responsible for detecting odors~\cite{Doty2001}.
An increased airflow penetration in this area at higher \textsc{Reynolds} numbers is likely to also raise the delivery of odorant molecules to these receptors which enhances olfactory perception.
Such a mechanism is consistent with the physiological role of sniffing, 
where accelerated airflow and the development of vortical structures improve odor sampling efficiency.

\begin{figure}
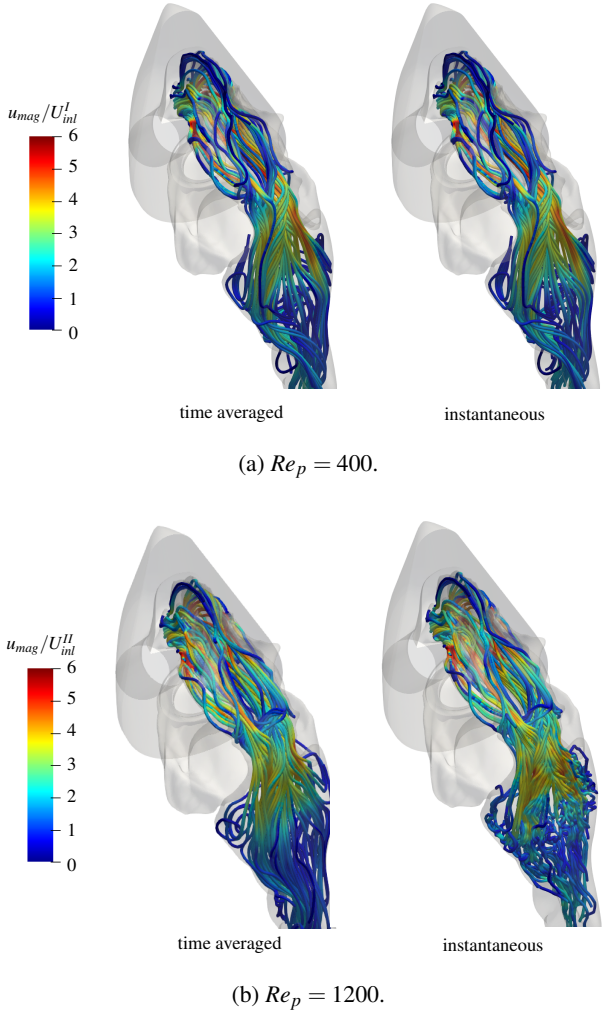

\captionsetup[subfigure]{aboveskip=8pt,belowskip=8pt}
\begin{subfigure}{\linewidth}
\centering
\include{figures/np_400}
\caption{$Re_p=400$.}
\label{fig:np_stream_400}
\end{subfigure}
\begin{subfigure}{\linewidth}
\centering
\include{figures/np_1200}
\caption{$Re_p=1200$.}
\label{fig:np_stream_1200}
\end{subfigure}
\caption{Streamlines of the time averaged and instantaneous velocity fields at $Re_p=400$ and $Re_p=1200$ colored by the normalized velocity magnitude $u_{mag}/U_{inl}$ in the nasopharynx and oropharynx.}
\label{fig:np}
\end{figure}

The second key anatomical region covers the nasopharynx and oropharynx.
Figure~\ref{fig:np} shows streamlines of the time averaged and instantaneous flow fields in this region, again colored by $u_{mag}/U_{inl}$.
The time averaged and instantaneous streamlines indicate the development of vortex pairs at $Re_p=400$ and $Re_p=1200$.
These counter‑rotating secondary flow structures are generated in the curved geometry due to centrifugal forces that shift higher‑momentum fluid towards the outer bend~\cite{Dean1959}.

At $Re_p=400$, the two vortices remain visually well-defined. 
The higher \textsc{Reynolds} number flow shows vortex pairs that increasingly overlap, 
especially further downstream, 
with pronounced differences in the time averaged and instantaneous streamlines.
In the curved section from the nasal cavity to the nasopharynx, 
the flow accelerates to normalized velocities up to $5$ times higher than $U_{inl}^I$. 
This enhances secondary vortex formation at both \textsc{Reynolds} numbers.
However, at higher $Re_p$, secondary instabilities are amplified which increases mixing of the vortical structures when the fluid enters the oropharynx.
Further downstream, the flow decelerates significantly due to the wider cross section which reduces the local strength of the vortices.
At $Re_p=1200$, this emphasizes the difference between the time averaged and instantaneous streamlines.
The latter show a more perturbed behavior that is characteristic of enhanced convective mixing.

\begin{figure}
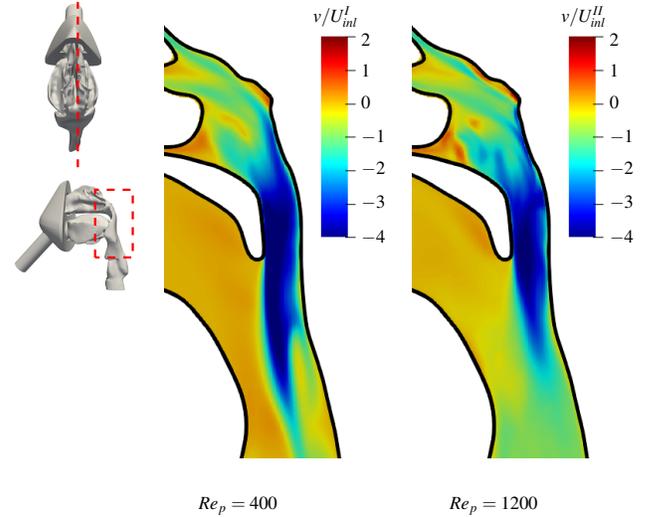

\centering
\include{figures/np_slice}
\caption{Time averaged normalized $y$-component of the velocity $v/U_{inl}$ at $Re_p=400$ and $Re_p=1200$ in a cross section in the right nasal passage (dashed red square and dashed red line in the sketch).}
\label{fig:np_slice}
\end{figure}

This trend is further illustrated in Fig.~\ref{fig:np_slice}, 
which shows the time averaged normalized $y$-velocity component $v$ in the cross section in the right nasal passage that is illustrated in the sketch.
At $Re_p=400$, the vertical jet remains coherent and expands downward with a relatively high velocity, 
which indicates limited cross-stream mixing.
In contrast, at $Re_p=1200$, the oropharyngeal jet is wider due to the more intense mixing.

\begin{figure}
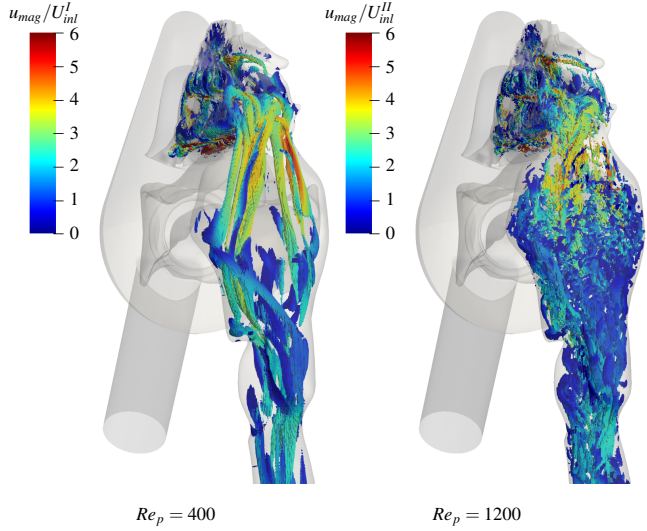

\centering
\include{figures/np_q}
\caption{$Q$-criterion of the instantaneous velocity fields at $Re_p=400$ and $Re_p=1200$ colored by the normalized velocity magnitude $u_{mag}/U_{inl}$ in the nasopharynx and oropharynx.}
\label{fig:np_q}
\end{figure}

To further investigate the secondary flow structures, 
the $Q$-criterion~\cite{Dubief00} is evaluated for the instantaneous velocity fields at $Re_p=400$ and $Re_p=1200$ in Fig.~\ref{fig:np_q}.
The quantity $Q$ is defined by
\begin{equation}
Q = \frac{1}{2}(\Omega_{jk}\Omega_{jk}-\Psi_{jk}\Psi_{jk}),
\end{equation}
where $\Psi_{jk}$ is the symmetric rate-of-strain tensor, $\Omega_{jk}$ is the anti-symmetric rate-of-rotation tensor,
and $j,k = 1, 2, 3$ are spatial coordinate indices in the $x$-, $y$- and $z$-directions.
\begin{figure}
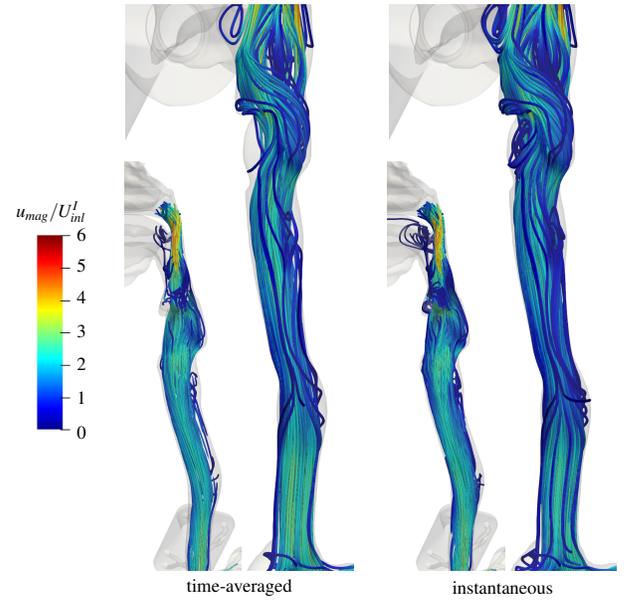
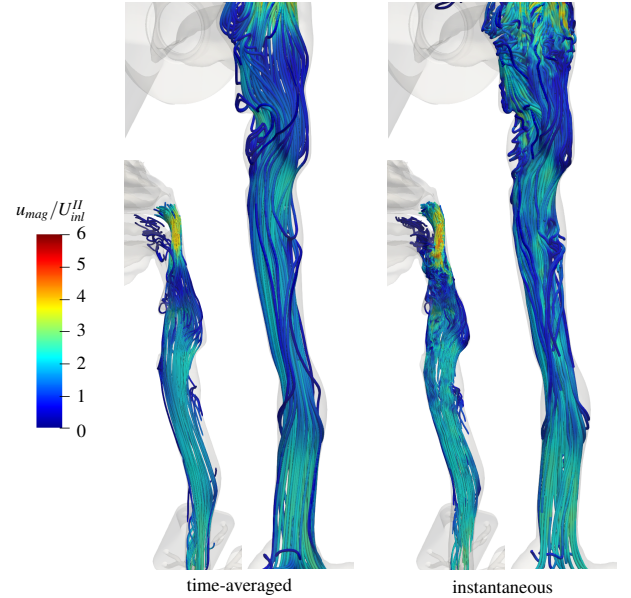

\captionsetup[subfigure]{aboveskip=8pt,belowskip=8pt}
\begin{subfigure}{\linewidth}
\centering
\include{figures/lp_400}
\caption{$Re_p=400$.}
\label{fig:lp_stream_400}
\end{subfigure}
\begin{subfigure}{\linewidth}
\centering
\include{figures/lp_1200}
\caption{$Re_p=1200$.}
\label{fig:lp_stream_1200}
\end{subfigure}
\caption{Streamlines of the time averaged and instantaneous velocity fields at $Re_p=400$ and $Re_p=1200$ colored by the normalized velocity magnitude $u_{mag}/U_{inl}$ in the laryngopharynx and larynx. For each case, the larger illustration presents the jet streamlines from an external perspective, while the smaller illustration provides an internal view with streamlines clipped in the symmetry plane of the $x$-axis, showing the core of the jet.}
\label{fig:lp_stream}
\end{figure}

At $Re_p=400$, coherent and stable vortex structures are clearly identifiable, particularly in the nasopharyngeal region, where the vortex pairs persist over time.
These structures are well-defined and remain localized, indicating a laminar flow regime with stable vortical structures.
In contrast, at $Re_p=1200$ the $Q$-criterion highlights the disintegration of large-scale structures into cascades of smaller, irregular vortices.
This breakdown is most prominent in the region downstream of the bend, where the flow decelerates and the mixing is intensified.
The loss of coherent vortical structures coincides with a massive growth of fine-scale vortices, 
resulting in increased spatial complexity and more disordered flow patterns.
This observation is consistent with the previously described streamline behavior, 
where the transition from organized flow in the bend to more chaotic flow in the oropharynx meant enhanced mixing.
The generation of small-scale vortical structures in this zone contributes to the dispersive transport of momentum.
It emphasizes the role of secondary instabilities in driving mixing at higher \textsc{Reynolds} numbers.

The third key anatomical region is the laryngopharynx and larynx. 
It contains the glottal jet, which originates at the glottis, i.e., the opening between the vocal folds in the larynx, and which is a key anatomical feature of respiration and phonation. 
As air passes through this constricted region, it forms a high-velocity core surrounded by pronounced shear layers near the walls. 
Physiological and in-vitro studies have extensively characterized this jet.
The results emphasize the significant influence of the strong flow acceleration and shear-layer formation on the downstream fluid dynamics~\cite{Krebs2011,Shinwari2003}.

Figure~\ref{fig:lp_stream} shows the glottal jet at $Re_p=400$ and $Re_p=1200$ by time averaged and instantaneous streamlines.
The larger illustrations present the jet from an external perspective, 
while the smaller illustrations provide an internal view on the symmetry plane of the $x$-axis, highlighting the core of the jet.
For $Re_p=400$, the time averaged and instantaneous streamlines agree closely.
That is, a clear jet core with confined shear layers is evident, 
especially in the symmetry plane. 
For $Re_p=1200$, the jet appears more diffuse and less confined,
since the radial momentum is increased, 
resulting in enhanced mixing.
This difference is more evident in the instantaneous streamlines.
Nevertheless, it can be stated that unlike in the nasopharynx and oropharynx regions, 
where time averaged and instantaneous streamlines deviated significantly at $Re_p=1200$, 
they remain similar in the laryngopharynx and larynx region.
\begin{figure}
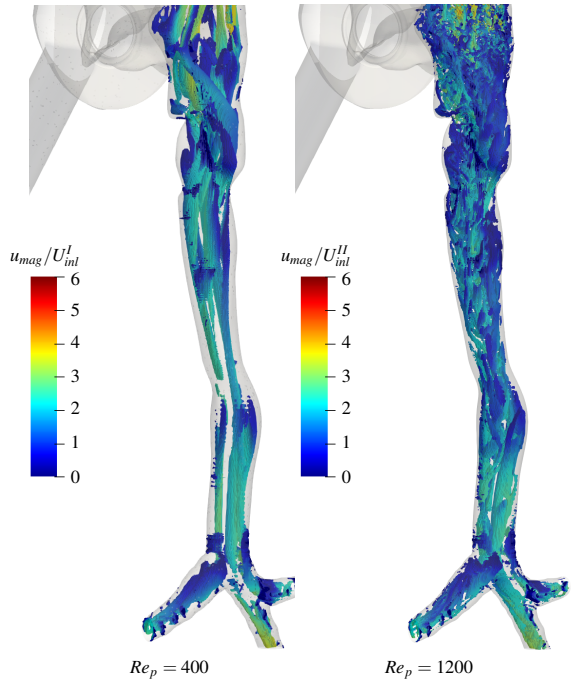

\centering
\include{figures/lp_q}
\caption{$Q$-Criterion of the instantaneous velocity fields at $Re_p=400$ and $Re_p=1200$ colored by the normalized velocity magnitude $u_{mag}/U_{inl}$ in the laryngopharynx and larynx.}
\label{fig:lp_q}
\end{figure}

The difference between the jets at $Re_p=400$ and $Re_p=1200$ becomes more obvious by analyzing the $Q$-criterion of the instantaneous flow fields in Fig.~\ref{fig:lp_q}.
At $Re_p=400$, a coherent, high-speed jet core surrounded by shear layers is visible. 
These shear layers roll up into streamwise vortices which appear as distinct pairs in the $Q$-criterion field. 
This evolution of the shear layer appears to be relatively stable which is typical for laminar or slightly transitional flows. 
The $Q$-distributions for $Re_p=1200$ show a clear deviation from an organized flow. 
The shear layers become unstable and develop irregular vortical structures. 
This indicates a highly unsteady flow which is characterized by complex vortex interactions.
The transition from symmetric to disordered, multiscale vortices evidences the sensitivity of the structure of the glottal jet to the \textsc{Reynolds} number and the increased mixing and shear-layer thickening at higher flow rates.

Further investigation of the distribution of high-speed regions along the airway centerline reveals that for both \textsc{Reynolds} numbers $u_{mag}$ exceeds $4 U_{inl}$. 
This is primarily true within the nasal cavity and the nasopharyngeal region in Figs.~\ref{fig:nc_streamlines} and~\ref{fig:np}. 
These regions act as geometric constrictions that locally accelerate the flow, 
particularly near the curved transitions and narrow passages.
Such geometric constrictions, e.g., caused by the nasal turbinates and valve regions, are crucial for heat and moisture exchange and thereby enhance mucosal contact with inhaled air~\cite{Zwicker2017}.
Temperature increase and humidification in the nasal cavity are strongly correlated and happen almost synchronously~\cite{Keck2001}.
However, narrow nasal passages also increase the airway resistance. 
To improve the breathing conditions by, e.g., nasal cavity surgeries, a compromise between reducing the pressure loss by widening and better conditioning the incoming air before entering the lung by narrowing the airway has to be found~\cite{Ruettgers2024,Ruettgers2025,Ruettgers2025a}.  %
In contrast to the nasal cavity or nasopharynx, 
in regions downstream from the oropharynx to the glottal jet lower peak velocities are observed, 
with local velocity magnitude values generally remaining below $4 U_{inl}$ (Figs.~\ref{fig:lp_stream} and~\ref{fig:lp_q}). 
This decrease correlates with the anatomical widening of the airway, 
which results in flow deceleration and redistribution.

\begin{figure}
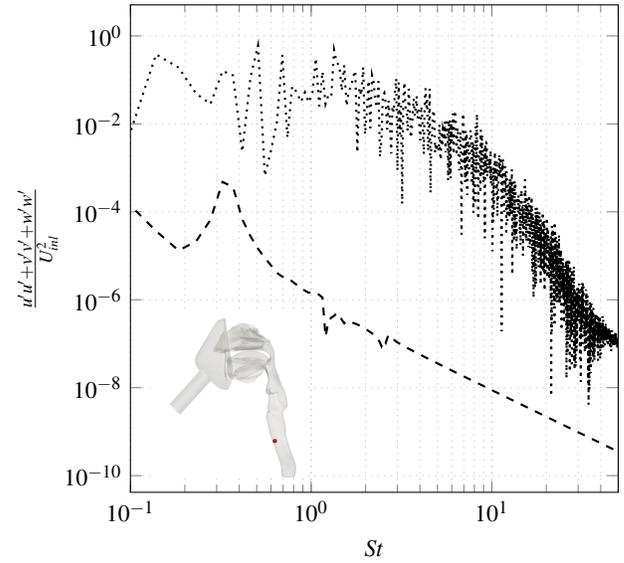

\centering
\include{figures/energy_spectra}
\caption{Velocity energy spectra in the shear layers of the glottal jet at the location illustrated by the red dot at $Re_p=400$ (dashed) and $Re_p=1200$ (dotted).}
\label{fig:energy}
\end{figure}

To complement the spatial analysis of the glottal jet and gain insight into the temporal dynamics of the shear layers, 
the energy spectrum of the velocity fluctuations at a probe point located within the shear region of the jet is evaluated in Fig.~\ref{fig:energy}.
The spectral energy $0.5 \cdot (u'u'+v'v'+w'w')$ normalized by $0.5 \cdot U_{inl}^2$ is shown as a function of the \textsc{Strouhal} number $St=fd_{inl}/U_{inl}$.
While the $Q$-criterion provides a snapshot of vortical structures at a given time, 
the energy spectrum captures the frequency content of the flow over time. 
This allows the assessment of the presence of coherent structures, dominant oscillatory modes, and broadband turbulence. 
In particular, the comparison of the spectra at $Re_p=400$ and $Re_p=1200$ yields a quantitative measure of how the shear-layer dynamics evolve from organized vortex roll-up to increasingly irregular, multi-frequency interactions.

At $Re_p=400$, the energy spectrum exhibits a clear peak at approximately $St=0.35$. 
This indicates the presence of the dominant large-scale, quasi-periodic structure which is likely associated with coherent shear-layer dynamics. 
Beyond this peak, the spectrum shows an almost linear decay in $\log-\log$ space. 
This suggests that the flow remains largely organized with only limited energy transfer to smaller turbulent scales.
\begin{figure}
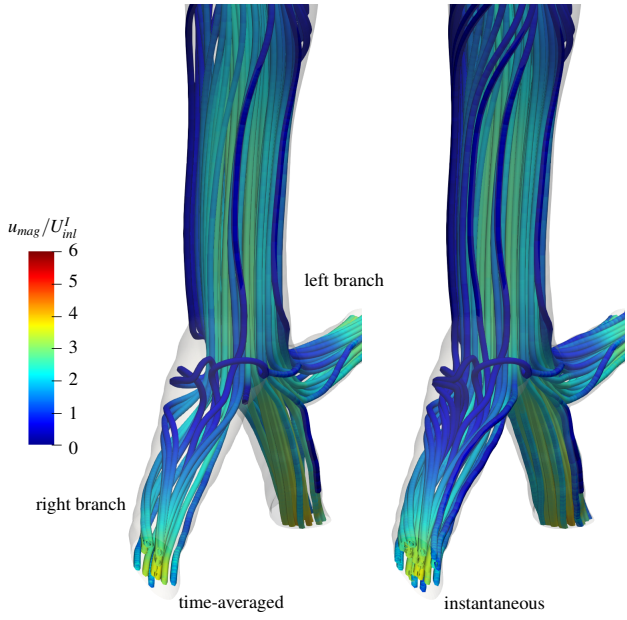
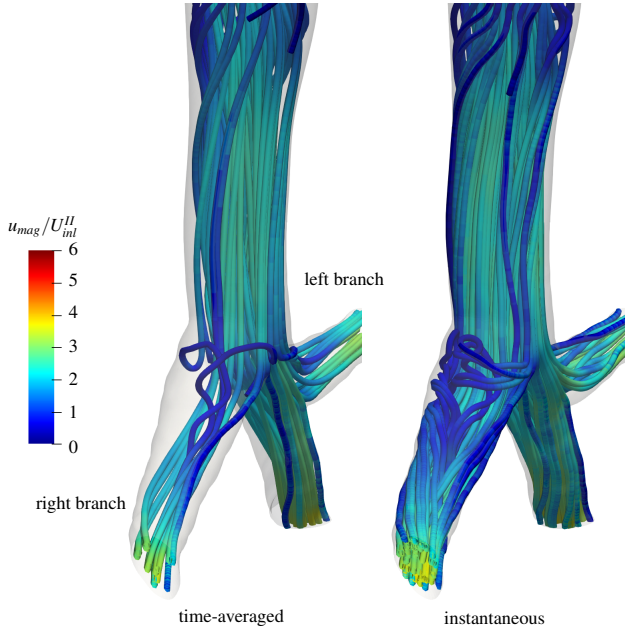

\captionsetup[subfigure]{aboveskip=8pt,belowskip=8pt}
\begin{subfigure}{\linewidth}
\centering
\include{figures/t_400}
\caption{$Re_p=400$.}
\label{fig:t_stream_400}
\end{subfigure}
\begin{subfigure}{\linewidth}
\centering
\include{figures/t_1200}
\caption{$Re_p=1200$.}
\label{fig:t_stream_1200}
\end{subfigure}
\caption{Streamlines of the time averaged and instantaneous velocity fields for $Re_p=400$ and $Re_p=1200$ colored by the normalized velocity magnitude $u_{mag}/U_{inl}$ in the tracheal and post-carinal regions.}
\label{fig:t_stream}
\end{figure}

For $Re_p=1200$, the spectrum reveals a more complex multi-regime structure.
A low-frequency range below $St=2$ captures the energy containing eddies and large-scale flow instabilities. 
This is followed by a well-defined intermediate region ranging from $St=2$ to $St=10$.
In this regime, the energy is transferred from larger to smaller scales without significant loss. 
At frequencies higher than $St=10$, the spectrum enters a dissipation range, where the energy content decays rapidly due to viscous effects.

The final key anatomical region contains the trachea, the carinal bifurcation, and the post-carinal region.
As can be seen in Fig.~\ref{fig:t_stream}, the time averaged and instantaneous streamlines at $Re_p=400$ and $Re_p=1200$ are largely aligned indicating limited temporal variability and a relatively stable flow regime. 
This result is similar to the glottal jet region. 
However, a key distinction emerges in the tracheal jet upstream of the carinal bifurcation. 
At $Re_p=400$, the jet remains well-defined with clearly visible shear layers surrounding the high-speed core. 
In contrast, at $Re_p=1200$ the jet is again more diffuse with broadened shear zones and reduced central velocity.
Again, this is due to enhanced momentum exchange and mixing upstream of the bifurcation.
\begin{figure}
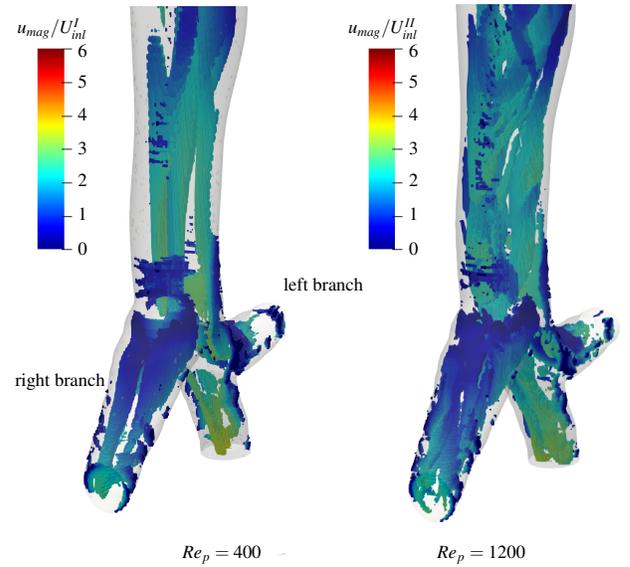

\centering
\include{figures/t_q}
\caption{$Q$-criterion of the instantaneous velocity fields at $Re_p=400$ and $Re_p=1200$ colored by the normalized velocity magnitude $u_{mag}/U_{inl}$ in the tracheal and post-carinal regions.}
\label{fig:t_q}
\end{figure}

Downstream of the bifurcation, these differences are less pronounced.
As discussed in Sec.~\ref{sec:res_1},
Figs.~\ref{fig:comparison_400} and~\ref{fig:comparison_1200} show a clear flow separation region and thin wall jet in the right branch
and a much smaller separation region with two wall jets that merge into one jet downstream in the left branch.
In the right branch, the flow patterns for both \textsc{Reynolds} numbers are characterized by separation and vortex pairs which differ quantitatively.
Overall, the geometry induced secondary flows dominate the post-bifurcation dynamics and impose a consistent structure on the downstream flow 
despite the upstream variations in the jet structure and turbulence intensity.
\begin{figure*}
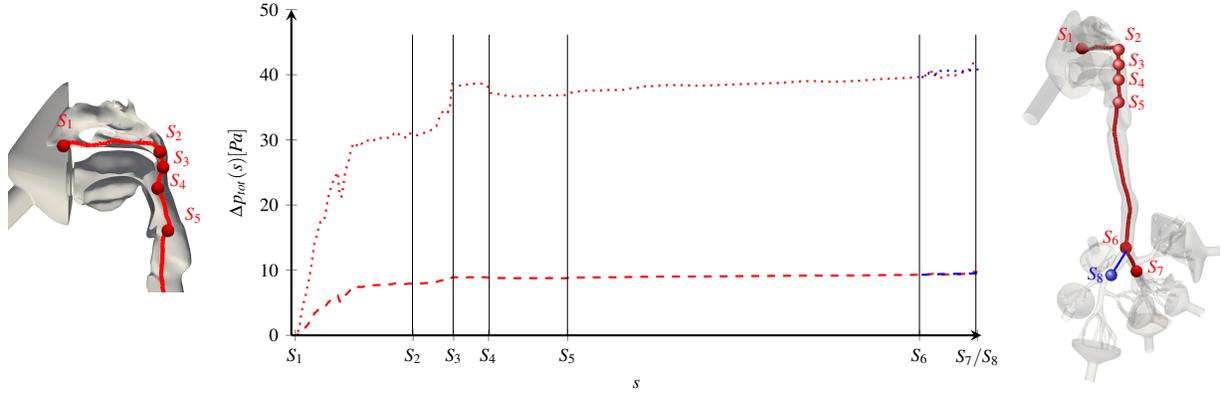

\centering
\include{figures/pressure_loss}
\caption{Total pressure loss between the right nostril and several locations along the airway beyond the carinal bifurcation for the $Re_p=400$ (dashed) and $Re_p=1200$ (dotted) cases. The various locations are indicated in the 3D models. The distributions between $S_6$ and $S_7$/$S_6$ and $S_8$ are shown by red/blue lines.}
\label{fig:del_p}
\end{figure*}

When the high-speed tracheal jet impinges upon the carinal bifurcation, 
it is divided into ducts with a smaller cross-sectional area relative to the wide glottal region. 
However, the cumulative cross-sectional area of the daughter branches exceeds that of the tracheal inlet resulting in an overall reduction in momentum and convective acceleration. 
This overall geometric expansion combined with the redistribution of the flow into curved ducts reduces asymmetries from the upstream flow. 
This development is observed for both \textsc{Reynolds} number flows.

This discussion is further supported by an analysis of the $Q$-criterion distributions of the instantaneous velocity field in Fig.~\ref{fig:t_q}.
For both \textsc{Reynolds} numbers, vortex pairs determine the flow in the bifurcated branches which lead to increased mixing and as such to a uniform velocity distribution further downstream.

\subsection{Pressure loss along the four key anatomical regions}
\label{sec:res_3}

The previously described flow features at the two \textsc{Reynolds} numbers affect the pressure drop, a key quantity in respiratory flows.
Elevated pressure loss along the airway increases the load on the respiratory muscles and can lead to reduced ventilation efficiency~\cite{Loring2009,Lintermann2013}. 
The pressure drop is extracted from the simulation results using the reference quantities for air, denoted in Tab.~\ref{tab:flow_param}.
The distribution of the time averaged total pressure $p_{tot}=p_{stat}+p_{dyn}$, with the dynamic pressure $p_{dyn} = \rho u_{mag}^2/2$, is determined along the geometrical centerline between the characteristic locations $S_1$ (right nostril) and $S_7$ / $S_8$ (post-carinal region).
Figure~\ref{fig:del_p} shows the distributions of the total pressure loss $\Delta p_{tot}(s) = \hat{p}_{tot}(S_1)-\hat{p}_{tot}(s)$ for both \textsc{Reynolds} numbers,
where $\hat{p}_{tot}(S_1)$ is the area averaged total pressure in the cross section of $S_1$
and $\hat{p}_{tot}(s)$ is the area averaged total pressure at the downstream cross section at location $s$ along the centerline. 
The normal vector of a cross section is computed based on two consecutive locations along the centerline.
The area of a cross section is determined by a region growing algorithm whose seed point is located at $s$.
The centerline is computed using the \textit{Vascular Modeling Toolkit} (VMTK)~\cite{Antiga2008}.

\begin{table}[]
\centering
\begin{tabular}{c|c|c|c} 
Key anatomical region & Points along & Change in & Contribution \\ [0.5ex]
 & centerline & $\Delta p_{tot} [Pa]$ & to $\Delta p_{tot}(S_7)$ \\
\hline
Nasal cavity & $S_1$ to $S_2$ & $+7.93$ & $81.25\%$ \\
Naso- and oropharynx & $S_2$ to $S_5$ & $+0.86$ & $8.81\%$ \\
Laryngopharynx and larynx & $S_5$ to $S_6$ & $+0.53$ & $5.43\%$ \\
Trachea and left car. branch & $S_6$ to $S_7$ & $+0.44$ & $4.51\%$ \\
Trachea and right car. branch & $S_6$ to $S_8$ & $+0.23$ & ---
\end{tabular}
\vspace*{+2mm}
\caption{Contributions of the four key anatomical regions to the total pressure loss for flow at $Re_p=400$.}
\label{tab:del_p_400}
\end{table}
\begin{table}[]
\centering
\begin{tabular}{c|c|c|c} 
Key anatomical region & Points along & Change in & Contribution \\ [0.5ex]
 & centerline & $\Delta p_{tot} [Pa]$ & to $\Delta p_{tot}(S_7)$ \\
\hline
Nasal cavity & $S_1$ to $S_2$ & $+30.76$ & $73.36\%$ \\
Naso- and oropharynx & $S_2$ to $S_5$ & $+6.23$ & $14.86\%$ \\
Laryngopharynx and larynx & $S_5$ to $S_6$ & $+2.57$ & $6.13\%$ \\
Trachea and left car. branch & $S_6$ to $S_7$ & $+2.37$ & $5.65\%$ \\
Trachea and right car. branch & $S_6$ to $S_8$ & $+1.19$ & ---
\end{tabular}
\vspace*{+2mm}
\caption{Contributions of the four key anatomical regions to the total pressure loss for flow at $Re_p=1200$.}
\label{tab:del_p_1200}
\end{table}

The pressure loss distributions between the right nostril ($S_1$) and seven characteristic points $S_2, \ldots,S_8$ along the centerline are analyzed.
The total pressure loss between $S_1$ and $S_7$ in the left post-carinal branch is $\Delta p_{tot}(S_7)=9.76~Pa$ for $Re_p=400$ and $\Delta p_{tot}(S_7)=41.93~Pa$ for $Re_p=1200$. 
That is, the latter is $4.3$ times larger than the former.
The contributions of each key anatomical region to $\Delta p_{tot}(S_7)$ are listed in Tabs.~\ref{tab:del_p_400} and~\ref{tab:del_p_1200}.
For both \textsc{Reynolds} numbers, the largest contributions occur in the nasal cavity between $S_1$ and $S_2$.
The pressure drop is caused by the narrow nasal valve region and the complex internal geometry defined by curved walls and turbinates. 
These geometric features cause the local flow to separate and enhance viscous dissipation. 
This is consistent with the velocity field observations presented in Sec.~\ref{sec:res_2}.
Although the streamline fields look similar for both \textsc{Reynolds} numbers, 
the higher normalized velocities at $Re_p=1200$, 
especially around the turbinates, 
mean higher kinetic energy and thus increased dynamic pressure.
The kink midway between $S_1$ and $S_2$ occurs due to the widening between cross sections III and IV in Fig.~\ref{fig:nc_cross_sec},
which has been discussed in Sec.~\ref{sec:res_2}.

The second largest difference in the pressure loss for $Re_p=400$ and $Re_p=1200$ is observed in the naso- and oropharyngeal regions between $S_2$ and $S_5$.
The pressure loss $\Delta p_{tot}$ increases by $0.86~Pa$ for $Re_p=400$ and by $6.23~Pa$ for $Re_p=1200$.
This increase is mainly due to the nasopharyngeal bend 
by which the flow is directed sharply downward. 
At $Re_p=400$, the flow remains laminar and the centrifugal forces generate vortex pairs, as shown in Fig.~\ref{fig:np_stream_400}. 
In this laminar flow, momentum is efficiently redistributed with limited energy dissipation, leading to only a mild pressure drop.
At $Re_p=1200$, however, the vortices disintegrate into a cascade of smaller vortical structures as evidenced by the $Q$-criterion analysis in Fig.~\ref{fig:np_q}. 
This breakdown of coherent motion and the development of fine-scale flow features enhances viscous dissipation and mixing, and significantly increases the total pressure loss.
A nearly constant pressure loss is observed in the narrow part of the oropharyngeal region between $S_3$ and $S_4$ regardless of the \textsc{Reynolds} number.
An expansion of the airway in the widened part of the oropharyngeal region between $S_4$ and $S_5$ reduces the pressure loss slightly in both flow regimes.

A further pressure loss increase is observed in the laryngopharynx and larynx, i.e., the region of the glottis jet between $S_5$ and $S_6$.
At $Re_p=400$, $\Delta p_{tot}$ increases by $0.53~Pa$ and by $2.57~Pa$ for $Re_p=1200$.
The gradient of the total pressure loss between $S_5$ and $S_6$ is small compared to that of the nasal cavity and nasopharyngeal bend for both \textsc{Reynolds} numbers.
This can be attributed to the distinct flow dynamics of the glottal jet.
While this region features high local velocities due to the constricted glottis, the jet is highly directed and essentially independent from strong wall interactions in its core.
At $Re_p=400$, the jet retains a coherent laminar structure with symmetric shear layers resulting in limited energy dissipation and thus a small total pressure drop of less than $1~Pa$.
At $Re_p=1200$, the total pressure loss is at a much higher level.
The jet is more diffuse and the shear layers break down into smaller vortices as evidenced by the $Q$-criterion and energy spectrum in Figs.~\ref{fig:lp_q} and~\ref{fig:energy}, which leads to enhanced mixing and the higher pressure loss level.
The geometric confinement of the glottis primarily accelerates the flow without inducing significant frictional losses along the airway wall such that the overall pressure gradient remains moderate.

In the final key anatomical region, the total pressure loss between the trachea ($S_6$) and the left ($S_7$) and right ($S_8$) branches of the carinal bifurcation are analyzed.
Between $S_6$ and $S_7$,
the pressure loss increases moderately by $2.37~Pa$ at $Re_p=1200$,
whereas almost no increase is observed at $Re_p=400$.
At $Re_p=400$, the jet remains focused and coherent with minimal shear-induced dissipation, leading to a negligible pressure drop in this segment.
In contrast, at $Re_p=1200$, the jet is already more diffuse upstream of the bifurcation, see Fig.~\ref{fig:t_stream_1200}, and the enhanced mixing within its broadened shear layers contributes to increased viscous losses as the flow transitions into the left daughter branch.

\begin{figure}
\centering
\include{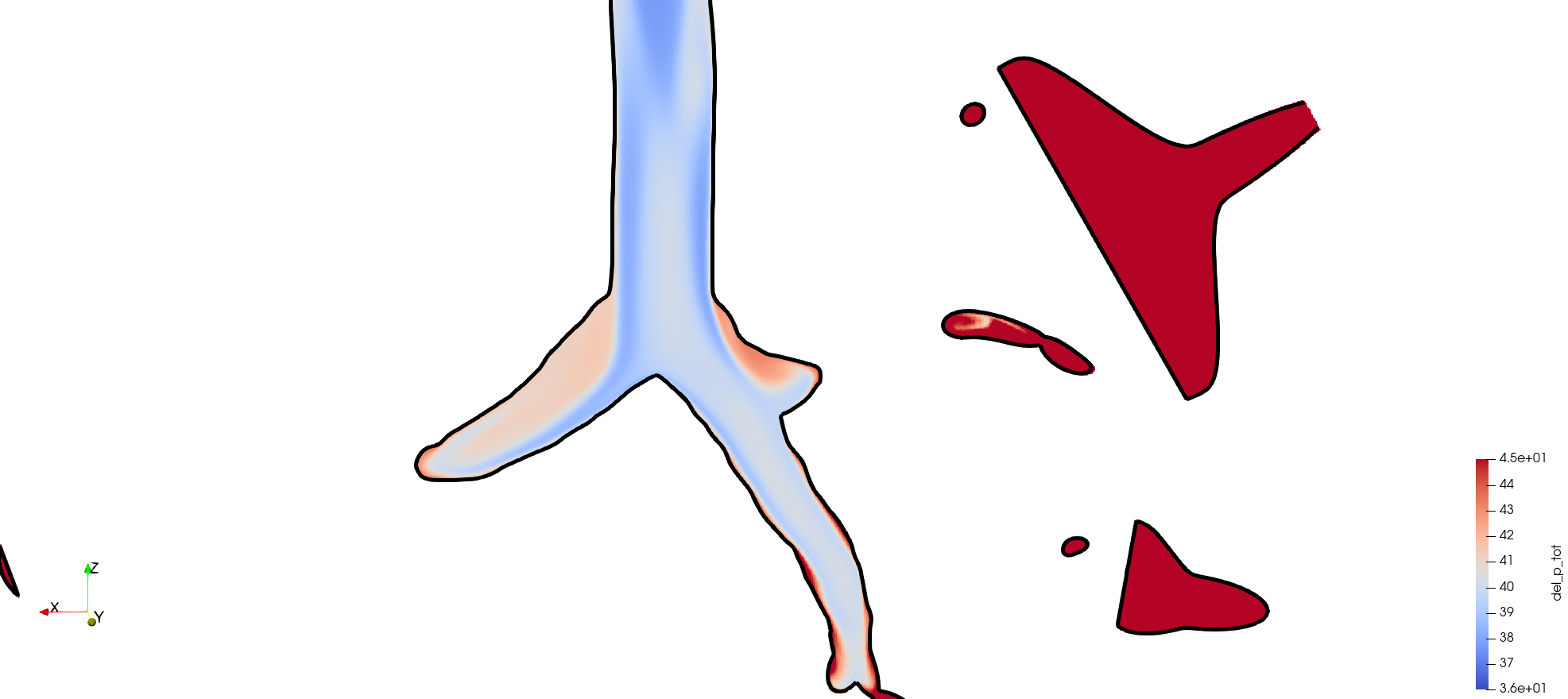}
\caption{Total pressure loss between the right nostril and the carinal bifurcation region for $Re_p=1200$. The location of the cross section is shown in the sketch of Fig.~\ref{fig:comparison_400}.}
\label{fig:del_p_bifu}
\end{figure}

Similar to the difference between $S_6$ and $S_7$, almost no increase in the total pressure is observed between $S_6$ and $S_8$ for $Re_p=400$.
For $Re_p=1200$, the increase is $1.19~Pa$.
That is, it is smaller than the increase between $S_6$ and $S_7$. 
Despite the larger recirculation zone in the right branch, as shown in Fig.~\ref{fig:comparison_1200}, 
Fig.~\ref{fig:del_p_bifu} indicates that the left branch exhibits a greater total pressure loss.
The increased loss in the left branch is caused by another bifurcation shortly downstream of the carinal bifurcation.
This introduces additional curvature, branching, and cross-sectional variation 
by which secondary flow structures are intensified.
They are additional sources of energy dissipation, particularly at the higher \textsc{Reynolds} number.

Nonetheless, the overall pressure loss remains modest compared to upstream regions.
This is likely due to the flow-stabilizing effect of the carinal bifurcation which has been discussed in Sec.~\ref{sec:res_1}.
The cumulative cross-sectional area of the daughter branches exceeds that of the tracheal inlet, resulting in an overall reduction in flow velocity and thus, in momentum and convective acceleration.

The relative contribution of the nasal cavity to the total pressure loss is with $81.25\%$ at $Re_p=400$ larger than $73.36\%$ at $Re_p=1200$. 
Interestingly, 
this is reversed for the naso- and oropharynx with $8.81\%$ at $Re_p=400$ and $14.86\%$ at $Re_p=1200$,
the laryngopharynx and larynx with $5.43\%$ at $Re_p=400$ and $6.13\%$ at $Re_p=1200$,
and the trachea and carinal bifurcation with $4.51\%$ and $5.65\%$ at $Re_p=400$  and $Re_p=1200$.

\section{Summary, discussion, and outlook}
\label{sec:disc}

Fully resolved, turbulence-model-free simulations of respiratory flows through a realistic human upper airway geometry at two physiologically relevant \textsc{Reynolds} numbers are presented. 
Previous numerical studies frequently relied on simplified geometries or were based on the \ac{RANS} equations plus a closure model such that the accuracy was limited among other constraints by modeling assumptions. 
To reduce the modeling impact, \ac{DNS} of nasal breathing in an airway model from a nasal mask to the 6th bronchial bifurcation were conducted.
The simulations were performed at $Re_p=400$ and $Re_p=1200$, representing resting and moderately elevated breathing conditions. 
A high-fidelity \ac{LB} method implemented in the open-source framework \ac{m-AIA} was validated against high-resolution \ac{PIV} and \ac{3D-PTV} measurements.

A comprehensive mesh refinement study was conducted to evaluate the impact of the spatial resolution on the accuracy of the numerical simulations. 
The coarse simulation underpredicted the jet core velocity and failed to resolve key flow structures observed in the measurements. 
The medium and fine meshes reproduced the spatial development of the jet for both \textsc{Reynolds} numbers. 
The fine mesh yielded the best agreement with the experimental data in velocity magnitude and flow structure. 
Therefore, the fine mesh was used for all subsequent analyses.

The flow field in four key anatomical regions of the airway was investigated: the nasal cavity, the naso- and oropharynx, the laryngopharynx and larynx (glottal jet region), and the trachea with the carinal bifurcation. 
In the following, the major findings of this study are summarized.

\begin{itemize}
    \setlength{\itemsep}{-0.2em}
    \item \textbf{Nasal cavity}: The nasal cavity exhibited the most significant total pressure loss in the entire airway, particularly at $Re_p=1200$, where $\Delta p_{tot}$ reached approximately $30.76~Pa$ compared to $7.93~Pa$ at $Re_p=400$.
    This is attributed to the narrow nasal valve and the complex internal geometry defined by the turbinates such that the flow is locally accelerated and vortices are generated.
    The analysis of the velocity field showed high-speed flow in the inferior and center turbinate channels.
    The streamline patterns remained similar between both \textsc{Reynolds} number flows. 
    The asymmetry between the left and right passages, which is likely linked to the nasal cycle, contributed to uneven pressure and velocity distributions that amplified local losses.
    At $Re_p=1200$, higher airflow into the uppermost part of the nasal cavity sensitized the neural receptors in the olfactory region. 
    \item \textbf{Nasopharynx and oropharynx}: The second notable total pressure loss occurred in the naso- and oropharyngeal regions, with $\Delta p_{tot}$ increasing by $6.23~Pa$ at $Re_p=1200$ and $1~Pa$ at $Re_p=400$.
    At the lower \textsc{Reynolds} number, secondary vortices were stable, resulting in a structured but low-dissipation flow.
    At $Re_p=1200$, vortices desintegrated into chaotic interactions and enhanced mixing.
    This increased dissipation led to higher total pressure loss.
    Further downstream, in the widened part of the oropharynx, the total pressure loss even slightly decreased.
    \item \textbf{Laryngopharynx and larynx}: In the glottal region, a further rise in the total pressure loss due to the different strength of the shear layers coating the jet was observed, i.e., $2.57~Pa$ at $Re_p=1200$ and less than $1~Pa$ at $Re_p=400$.
    In general, the glottis acts as a geometric nozzle that channels the flow and limits the overall pressure drop relative to more constricted upstream regions.
    \item \textbf{Trachea and carinal bifurcation}:
    This segment showed a moderate pressure rise of $2.37~Pa$ for $Re_p=1200$ and hardly any changing pressure at $Re_p=400$ for the flow from the trachea into the left branch downstream of the bifurcation.
    Similarly, for the flow into the right branch almost no change in pressure was observed at $Re_p=400$.
    However, the pressure difference of $1.19~Pa$ for $Re_p=1200$ was smaller than in the left branch.
    The increased total pressure loss in the left branch comes mainly from another bifurcation shortly downstream of the main carinal bifurcation.
    Nevertheless, the overall pressure loss remained modest compared to upstream regions.
    When the tracheal jet reaches the carinal bifurcation, it splits into narrower but cumulatively wider daughter branches, leading to reduced momentum and convective acceleration.
\end{itemize}

The present findings showed that the nasal cavity plays a key role in olfactory transport. 
Under normal resting conditions, only a small fraction of the inhaled air ($5-15\%$) reaches the olfactory region.
Most of the flow bypasses it along the lower part of the nasal cavity~\cite{Kelly2000}.
At higher \textsc{Reynolds} numbers, an increased flow penetration into the superior nasal regions was observed.
This corroborates the conclusions by Zhao et al.~\cite{Zhao2006} and Kelly et al.~\cite{Kelly2000}, 
who found that enhanced airflow towards the olfactory system improved sensory perception during sniffing. 
The results of the current study clearly agree with this interpretation and provide further confirmation by cross-sectional velocity field data.

The transition from stable secondary vortices at $Re_p=400$ to chaotic secondary flows at $Re_p=1200$ in the naso- and oropharyngeal regions leads to enhanced convective mixing. 
On the one hand, stronger mixing promotes more uniform heat and humidity exchange.
On the other hand, higher mixing also increases wall contact and residence times for inhaled particles and droplets resulting in adverse implications for:
\begin{itemize}
    \setlength{\itemsep}{-0.2em}
    \item Pathogen transmission: Turbulent eddies can carry infectious aerosols toward the mucosal surface, facilitating viral uptake~\cite{Asgharian2001}.
    \item Mucosal irritation: Inhalation of pollutants or irritants is more likely to result in contact with sensitive pharyngeal tissue.
    \item Uncontrolled drug deposition: For inhalation therapies targeting the lower airways, excessive upstream mixing increases the risk of premature deposition in the pharynx, which might reduce the therapeutic dose reaching distal regions~\cite{Zhang2001}.
\end{itemize}
These effects should be considered in the design of aerosol-based treatments and surgical interventions that alter pharyngeal geometry.

Regarding the glottal jet, at the lower \textsc{Reynolds} number, the jet had a focused core with coherent shear layers and symmetric roll-up into streamwise vortices. 
In contrast, the higher \textsc{Reynolds} number case showed widened shear layers and shear-layer breakdown that generated turbulence.
Similar patterns have been reported in laboratory models.
At higher \textsc{Reynolds} number, the glottal flow changes from laminar-like jets to turbulent shear layers that enhance mixing and shear-induced energy dissipation~\cite{Shinwari2003}.
For aerosol therapies targeting lower airways, the glottal transition influences aerosol trajectory and dispersion. 
Higher turbulence may improve downstream penetration, but also increases the deposition likelihood at the vocal fold region, potentially exposing the local tissue to irritants or drugs.

A novel and central contribution of the present study is the systematic discussion of the pressure loss in four anatomically distinct regions of the airway. 
This multi-regional analysis revealed that pressure loss is not only a passive byproduct of airway resistance, but also a sensitive marker for the transition from laminar flow to multiscale dynamics. 
Interestingly, while the nasal cavity accounts for the largest share of the absolute pressure loss at both flow rates, its relative contribution decreases from over $80\%$ at $Re_p = 400$ to about $73\%$ at $Re_p = 1200$.
This indicates that downstream regions contribute more significantly at higher flow rates.

Future work will extend these analyses to include comparisons between nasal and oral breathing modes at the current \textsc{Reynolds} numbers 
as well as simulations and measurements covering complete breathing cycles. 
Those investigations aim to provide an even more comprehensive understanding of patient-specific airflow patterns and to support the design of optimized inhalation therapies and surgical interventions tailored to individual anatomies.

\section*{Conflict of Interest statement}
This statement is to declare that the authors of this manuscript, Mario Rüttgers, Julian Vorspohl, Luca Mayolle, Benedikt Johanning-Meiners, Dominik Krug, Michael Klaas, Matthias Meinke, Sangseung Lee, Wolfgang Schröder, and Andreas Lintermann do not possess any financial dependence that might bias this work. The authors hereby declare that no conflict of interest exists in this work.

\begin{acknowledgments}
The research leading to these results has been funded by the German Research Foundation within the Walter Benjamin fellowship RU~2771/1-1 and by the HANAMI project within the European Union Horizon Europe Programme - Grant Agreement Number 101136269 under the call HORIZON-EUROHPC-JU-2022-INCO-04.
The authors gratefully acknowledge computing time on the supercomputer JURECA~\cite{JURECA} at Forschungszentrum Jülich under grant \textit{jhpc54}.
\end{acknowledgments}

\section*{Data Availability Statement}
The data that support the findings of this study are available from the corresponding author upon reasonable request.

\section*{Author contributions}
M.R. conceptualized the project, worked on the numerical methods, conducted the numerical simulations, analyzed the simulation and measurement results, and wrote the manuscript.
J.V. worked on the numerical methods, analyzed the simulation and measurement results, and edited the document.
L.M. and B.JM. conducted the experiments, analyzed the experimental results, and provided advise from the experimental fluid mechanics perspective.
D.K. and S.L. provided advise from a general fluid mechanics perspective.
W.S. provided advise from a general fluid mechanics perspective and edited the structure of the text and the discussion of the results.
M.M. and M.K. acquired funding.
A.L. acquired funding, supervised the project, provided advise from a general fluid mechanics perspectives, and edited the document.

\section*{References}

\nocite{*}
\bibliography{aipsamp}

\providecommand{\noopsort}[1]{}\providecommand{\singleletter}[1]{#1}%
\begin{thebibliography}{73}%
\makeatletter
\providecommand \@ifxundefined [1]{%
 \@ifx{#1\undefined}
}%
\providecommand \@ifnum [1]{%
 \ifnum #1\expandafter \@firstoftwo
 \else \expandafter \@secondoftwo
 \fi
}%
\providecommand \@ifx [1]{%
 \ifx #1\expandafter \@firstoftwo
 \else \expandafter \@secondoftwo
 \fi
}%
\providecommand \natexlab [1]{#1}%
\providecommand \enquote  [1]{``#1''}%
\providecommand \bibnamefont  [1]{#1}%
\providecommand \bibfnamefont [1]{#1}%
\providecommand \citenamefont [1]{#1}%
\providecommand \href@noop [0]{\@secondoftwo}%
\providecommand \href [0]{\begingroup \@sanitize@url \@href}%
\providecommand \@href[1]{\@@startlink{#1}\@@href}%
\providecommand \@@href[1]{\endgroup#1\@@endlink}%
\providecommand \@sanitize@url [0]{\catcode `\\12\catcode `\$12\catcode
  `\&12\catcode `\#12\catcode `\^12\catcode `\_12\catcode `\%12\relax}%
\providecommand \@@startlink[1]{}%
\providecommand \@@endlink[0]{}%
\providecommand \url  [0]{\begingroup\@sanitize@url \@url }%
\providecommand \@url [1]{\endgroup\@href {#1}{\urlprefix }}%
\providecommand \urlprefix  [0]{URL }%
\providecommand \Eprint [0]{\href }%
\providecommand \doibase [0]{http://dx.doi.org/}%
\providecommand \selectlanguage [0]{\@gobble}%
\providecommand \bibinfo  [0]{\@secondoftwo}%
\providecommand \bibfield  [0]{\@secondoftwo}%
\providecommand \translation [1]{[#1]}%
\providecommand \BibitemOpen [0]{}%
\providecommand \bibitemStop [0]{}%
\providecommand \bibitemNoStop [0]{.\EOS\space}%
\providecommand \EOS [0]{\spacefactor3000\relax}%
\providecommand \BibitemShut  [1]{\csname bibitem#1\endcsname}%
\let\auto@bib@innerbib\@empty
\bibitem [{\citenamefont {Dutta}\ \emph {et~al.}(2020)\citenamefont {Dutta},
  \citenamefont {Spence}, \citenamefont {Wei}, \citenamefont {Dhapare},
  \citenamefont {Hindle},\ and\ \citenamefont {Longest}}]{Dutta2020}%
  \BibitemOpen
  \bibfield  {author} {\bibinfo {author} {\bibfnamefont {R.}~\bibnamefont
  {Dutta}}, \bibinfo {author} {\bibfnamefont {B.}~\bibnamefont {Spence}},
  \bibinfo {author} {\bibfnamefont {X.}~\bibnamefont {Wei}}, \bibinfo {author}
  {\bibfnamefont {S.}~\bibnamefont {Dhapare}}, \bibinfo {author} {\bibfnamefont
  {M.}~\bibnamefont {Hindle}}, \ and\ \bibinfo {author} {\bibfnamefont
  {P.}~\bibnamefont {Longest}},\ }\bibfield  {title} {\enquote {\bibinfo
  {title} {Cfd guided optimization of nose-to-lung aerosol delivery in adults:
  Effects of inhalation waveforms and synchronized aerosol delivery},}\ }\href
  {\doibase 10.1007/s11095-020-02923-8} {\bibfield  {journal} {\bibinfo
  {journal} {Pharmaceutical research}\ }\textbf {\bibinfo {volume} {37}},\
  \bibinfo {pages} {199} (\bibinfo {year} {2020})}\BibitemShut {NoStop}%
\bibitem [{\citenamefont {Dastoorian}\ \emph {et~al.}(2022)\citenamefont
  {Dastoorian}, \citenamefont {Pakzad}, \citenamefont {Kozinski},\ and\
  \citenamefont {Behzadfar}}]{Dastoorian2022}%
  \BibitemOpen
  \bibfield  {author} {\bibinfo {author} {\bibfnamefont {F.}~\bibnamefont
  {Dastoorian}}, \bibinfo {author} {\bibfnamefont {L.}~\bibnamefont {Pakzad}},
  \bibinfo {author} {\bibfnamefont {J.}~\bibnamefont {Kozinski}}, \ and\
  \bibinfo {author} {\bibfnamefont {E.}~\bibnamefont {Behzadfar}},\ }\bibfield
  {title} {\enquote {\bibinfo {title} {A cfd investigation on the aerosol drug
  delivery in the mouth–throat airway using a pressurized metered-dose
  inhaler device},}\ }\href {\doibase 10.3390/pr10071230} {\bibfield  {journal}
  {\bibinfo  {journal} {Processes}\ }\textbf {\bibinfo {volume} {10}} (\bibinfo
  {year} {2022}),\ 10.3390/pr10071230}\BibitemShut {NoStop}%
\bibitem [{\citenamefont {Rüttgers}\ \emph {et~al.}(2024)\citenamefont
  {Rüttgers}, \citenamefont {Waldmann}, \citenamefont {Vogt}, \citenamefont
  {Ilgner}, \citenamefont {Schröder},\ and\ \citenamefont
  {Lintermann}}]{Ruettgers2024}%
  \BibitemOpen
  \bibfield  {author} {\bibinfo {author} {\bibfnamefont {M.}~\bibnamefont
  {Rüttgers}}, \bibinfo {author} {\bibfnamefont {M.}~\bibnamefont {Waldmann}},
  \bibinfo {author} {\bibfnamefont {K.}~\bibnamefont {Vogt}}, \bibinfo {author}
  {\bibfnamefont {J.}~\bibnamefont {Ilgner}}, \bibinfo {author} {\bibfnamefont
  {W.}~\bibnamefont {Schröder}}, \ and\ \bibinfo {author} {\bibfnamefont
  {A.}~\bibnamefont {Lintermann}},\ }\bibfield  {title} {\enquote {\bibinfo
  {title} {Automated surgery planning for an obstructed nose by combining
  computational fluid dynamics with reinforcement learning},}\ }\href {\doibase
  10.1016/j.compbiomed.2024.108383} {\bibfield  {journal} {\bibinfo  {journal}
  {Computers in Biology and Medicine}\ }\textbf {\bibinfo {volume} {173}},\
  \bibinfo {pages} {108383} (\bibinfo {year} {2024})}\BibitemShut {NoStop}%
\bibitem [{\citenamefont {Waldmann}\ \emph {et~al.}(2022)\citenamefont
  {Waldmann}, \citenamefont {Rüttgers}, \citenamefont {Lintermann},\ and\
  \citenamefont {Schröder}}]{Waldmann2022}%
  \BibitemOpen
  \bibfield  {author} {\bibinfo {author} {\bibfnamefont {M.}~\bibnamefont
  {Waldmann}}, \bibinfo {author} {\bibfnamefont {M.}~\bibnamefont {Rüttgers}},
  \bibinfo {author} {\bibfnamefont {A.}~\bibnamefont {Lintermann}}, \ and\
  \bibinfo {author} {\bibfnamefont {W.}~\bibnamefont {Schröder}},\ }\bibfield
  {title} {\enquote {\bibinfo {title} {Virtual surgeries of nasal cavities
  using a coupled lattice-boltzmann–level-set approach},}\ }\href {\doibase
  10.1115/1.4054042} {\bibfield  {journal} {\bibinfo  {journal} {Journal of
  Engineering and Science in Medical Diagnostics and Therapy}\ }\textbf
  {\bibinfo {volume} {5}} (\bibinfo {year} {2022}),\
  10.1115/1.4054042}\BibitemShut {NoStop}%
\bibitem [{\citenamefont {Carson}, \citenamefont {{Van Loon}},\ and\
  \citenamefont {Arora}(2024)}]{Carson2024}%
  \BibitemOpen
  \bibfield  {author} {\bibinfo {author} {\bibfnamefont {J.~M.}\ \bibnamefont
  {Carson}}, \bibinfo {author} {\bibfnamefont {R.}~\bibnamefont {{Van Loon}}},
  \ and\ \bibinfo {author} {\bibfnamefont {H.}~\bibnamefont {Arora}},\
  }\bibfield  {title} {\enquote {\bibinfo {title} {A personalised computational
  model of the impact of covid-19 on lung function under mechanical
  ventilation},}\ }\href {\doibase 10.1016/j.compbiomed.2024.109177} {\bibfield
   {journal} {\bibinfo  {journal} {Computers in Biology and Medicine}\ }\textbf
  {\bibinfo {volume} {183}},\ \bibinfo {pages} {109177} (\bibinfo {year}
  {2024})}\BibitemShut {NoStop}%
\bibitem [{\citenamefont {Lintermann}\ and\ \citenamefont
  {Schr{\"{o}}der}(2017)}]{Lintermann2017}%
  \BibitemOpen
  \bibfield  {author} {\bibinfo {author} {\bibfnamefont {A.}~\bibnamefont
  {Lintermann}}\ and\ \bibinfo {author} {\bibfnamefont {W.}~\bibnamefont
  {Schr{\"{o}}der}},\ }\bibfield  {title} {\enquote {\bibinfo {title}
  {{Simulation of aerosol particle deposition in the upper human
  tracheobronchial tract}},}\ }\href {\doibase
  10.1016/j.euromechflu.2017.01.008} {\bibfield  {journal} {\bibinfo  {journal}
  {European Journal of Mechanics - B/Fluids}\ }\textbf {\bibinfo {volume}
  {63}},\ \bibinfo {pages} {73--89} (\bibinfo {year} {2017})}\BibitemShut
  {NoStop}%
\bibitem [{\citenamefont {R{\"u}ttgers}\ \emph {et~al.}(2025)\citenamefont
  {R{\"u}ttgers}, \citenamefont {H{\"u}benthal}, \citenamefont {Tsubokura},\
  and\ \citenamefont {Lintermann}}]{Ruettgers2025a}%
  \BibitemOpen
  \bibfield  {author} {\bibinfo {author} {\bibfnamefont {M.}~\bibnamefont
  {R{\"u}ttgers}}, \bibinfo {author} {\bibfnamefont {F.}~\bibnamefont
  {H{\"u}benthal}}, \bibinfo {author} {\bibfnamefont {M.}~\bibnamefont
  {Tsubokura}}, \ and\ \bibinfo {author} {\bibfnamefont {A.}~\bibnamefont
  {Lintermann}},\ }\bibfield  {title} {\enquote {\bibinfo {title} {Parallel
  reinforcement learning and gaussian process regression for improved
  physics-based nasal surgery planning},}\ }in\ \href@noop {} {\emph {\bibinfo
  {booktitle} {Parallel Processing and Applied Mathematics}}},\ \bibinfo
  {editor} {edited by\ \bibinfo {editor} {\bibfnamefont {R.}~\bibnamefont
  {Wyrzykowski}}, \bibinfo {editor} {\bibfnamefont {J.}~\bibnamefont
  {Dongarra}}, \bibinfo {editor} {\bibfnamefont {E.}~\bibnamefont {Deelman}}, \
  and\ \bibinfo {editor} {\bibfnamefont {K.}~\bibnamefont {Karczewski}}}\
  (\bibinfo  {publisher} {Springer Nature Switzerland},\ \bibinfo {address}
  {Cham},\ \bibinfo {year} {2025})\ pp.\ \bibinfo {pages} {79--96}\BibitemShut
  {NoStop}%
\bibitem [{\citenamefont {Rüttgers}\ \emph {et~al.}(2026)\citenamefont
  {Rüttgers}, \citenamefont {Waldmann}, \citenamefont {Hübenthal},
  \citenamefont {Vogt}, \citenamefont {Tsubokura}, \citenamefont {Lee},\ and\
  \citenamefont {Lintermann}}]{Ruettgers2025}%
  \BibitemOpen
  \bibfield  {author} {\bibinfo {author} {\bibfnamefont {M.}~\bibnamefont
  {Rüttgers}}, \bibinfo {author} {\bibfnamefont {M.}~\bibnamefont {Waldmann}},
  \bibinfo {author} {\bibfnamefont {F.}~\bibnamefont {Hübenthal}}, \bibinfo
  {author} {\bibfnamefont {K.}~\bibnamefont {Vogt}}, \bibinfo {author}
  {\bibfnamefont {M.}~\bibnamefont {Tsubokura}}, \bibinfo {author}
  {\bibfnamefont {S.}~\bibnamefont {Lee}}, \ and\ \bibinfo {author}
  {\bibfnamefont {A.}~\bibnamefont {Lintermann}},\ }\bibfield  {title}
  {\enquote {\bibinfo {title} {Towards a widespread usage of computational
  fluid dynamics simulations for automated virtual nasal surgery planning},}\
  }\href {\doibase 10.1016/j.future.2025.107935} {\bibfield  {journal}
  {\bibinfo  {journal} {Future Generation Computer Systems}\ }\textbf {\bibinfo
  {volume} {174}},\ \bibinfo {pages} {107935} (\bibinfo {year}
  {2026})}\BibitemShut {NoStop}%
\bibitem [{\citenamefont {Lintermann}, \citenamefont {Meinke},\ and\
  \citenamefont {Schröder}(2013)}]{Lintermann2013}%
  \BibitemOpen
  \bibfield  {author} {\bibinfo {author} {\bibfnamefont {A.}~\bibnamefont
  {Lintermann}}, \bibinfo {author} {\bibfnamefont {M.}~\bibnamefont {Meinke}},
  \ and\ \bibinfo {author} {\bibfnamefont {W.}~\bibnamefont {Schröder}},\
  }\bibfield  {title} {\enquote {\bibinfo {title} {Fluid mechanics based
  classification of the respiratory efficiency of several nasal cavities},}\
  }\href {\doibase 10.1016/j.compbiomed.2013.09.003} {\bibfield  {journal}
  {\bibinfo  {journal} {Computers in Biology and Medicine}\ }\textbf {\bibinfo
  {volume} {43}},\ \bibinfo {pages} {1833--1852} (\bibinfo {year}
  {2013})}\BibitemShut {NoStop}%
\bibitem [{\citenamefont {Niegodajew}(2025)}]{Niegodajew2025}%
  \BibitemOpen
  \bibfield  {author} {\bibinfo {author} {\bibfnamefont {P.}~\bibnamefont
  {Niegodajew}},\ }\bibfield  {title} {\enquote {\bibinfo {title} {Flow
  patterns and vortex formation mechanisms inside a human nasal cavity},}\
  }\href {\doibase 10.1063/5.0253363} {\bibfield  {journal} {\bibinfo
  {journal} {Physics of Fluids}\ }\textbf {\bibinfo {volume} {37}},\ \bibinfo
  {pages} {021918} (\bibinfo {year} {2025})}\BibitemShut {NoStop}%
\bibitem [{\citenamefont {Emmerling}\ \emph {et~al.}(2024)\citenamefont
  {Emmerling}, \citenamefont {Vahaji}, \citenamefont {Morton}, \citenamefont
  {Fletcher},\ and\ \citenamefont {Inthavong}}]{Emmerling2024}%
  \BibitemOpen
  \bibfield  {author} {\bibinfo {author} {\bibfnamefont {J.}~\bibnamefont
  {Emmerling}}, \bibinfo {author} {\bibfnamefont {S.}~\bibnamefont {Vahaji}},
  \bibinfo {author} {\bibfnamefont {D.~A.}\ \bibnamefont {Morton}}, \bibinfo
  {author} {\bibfnamefont {D.~F.}\ \bibnamefont {Fletcher}}, \ and\ \bibinfo
  {author} {\bibfnamefont {K.}~\bibnamefont {Inthavong}},\ }\bibfield  {title}
  {\enquote {\bibinfo {title} {Scale resolving simulations of the effect of
  glottis motion and the laryngeal jet on flow dynamics during respiration},}\
  }\href {\doibase 10.1016/j.cmpb.2024.108064} {\bibfield  {journal} {\bibinfo
  {journal} {Computer Methods and Programs in Biomedicine}\ }\textbf {\bibinfo
  {volume} {247}},\ \bibinfo {pages} {108064} (\bibinfo {year}
  {2024})}\BibitemShut {NoStop}%
\bibitem [{\citenamefont {Yang}\ \emph {et~al.}(2020)\citenamefont {Yang},
  \citenamefont {Higano}, \citenamefont {Gunatilaka}, \citenamefont {Hysinger},
  \citenamefont {Amin}, \citenamefont {Woods},\ and\ \citenamefont
  {Bates}}]{Yang2020}%
  \BibitemOpen
  \bibfield  {author} {\bibinfo {author} {\bibfnamefont {M.}~\bibnamefont
  {Yang}}, \bibinfo {author} {\bibfnamefont {N.}~\bibnamefont {Higano}},
  \bibinfo {author} {\bibfnamefont {C.}~\bibnamefont {Gunatilaka}}, \bibinfo
  {author} {\bibfnamefont {E.}~\bibnamefont {Hysinger}}, \bibinfo {author}
  {\bibfnamefont {R.}~\bibnamefont {Amin}}, \bibinfo {author} {\bibfnamefont
  {J.}~\bibnamefont {Woods}}, \ and\ \bibinfo {author} {\bibfnamefont
  {A.}~\bibnamefont {Bates}},\ }\bibfield  {title} {\enquote {\bibinfo {title}
  {Subglottic stenosis position affects work of breathing},}\ }\href {\doibase
  10.1002/lary.29169} {\bibfield  {journal} {\bibinfo  {journal} {The
  Laryngoscope}\ }\textbf {\bibinfo {volume} {131}},\ \bibinfo {pages}
  {E1220--E1226} (\bibinfo {year} {2020})}\BibitemShut {NoStop}%
\bibitem [{\citenamefont {Xu}\ \emph {et~al.}(2020)\citenamefont {Xu},
  \citenamefont {Wu}, \citenamefont {Weng},\ and\ \citenamefont {Fu}}]{Xu2020}%
  \BibitemOpen
  \bibfield  {author} {\bibinfo {author} {\bibfnamefont {X.}~\bibnamefont
  {Xu}}, \bibinfo {author} {\bibfnamefont {J.}~\bibnamefont {Wu}}, \bibinfo
  {author} {\bibfnamefont {W.}~\bibnamefont {Weng}}, \ and\ \bibinfo {author}
  {\bibfnamefont {M.}~\bibnamefont {Fu}},\ }\bibfield  {title} {\enquote
  {\bibinfo {title} {Investigation of inhalation and exhalation flow pattern in
  a realistic human upper airway model by piv experiments and cfd
  simulations},}\ }\href {\doibase 10.1007/s10237-020-01299-3} {\bibfield
  {journal} {\bibinfo  {journal} {Biomechanics and Modeling in Mechanobiology}\
  }\textbf {\bibinfo {volume} {19}} (\bibinfo {year} {2020}),\
  10.1007/s10237-020-01299-3}\BibitemShut {NoStop}%
\bibitem [{\citenamefont {Morita}\ \emph {et~al.}(2022)\citenamefont {Morita},
  \citenamefont {Takeishi}, \citenamefont {Wada},\ and\ \citenamefont
  {Hatakeyama}}]{Morita2022}%
  \BibitemOpen
  \bibfield  {author} {\bibinfo {author} {\bibfnamefont {K.}~\bibnamefont
  {Morita}}, \bibinfo {author} {\bibfnamefont {N.}~\bibnamefont {Takeishi}},
  \bibinfo {author} {\bibfnamefont {S.}~\bibnamefont {Wada}}, \ and\ \bibinfo
  {author} {\bibfnamefont {T.}~\bibnamefont {Hatakeyama}},\ }\bibfield  {title}
  {\enquote {\bibinfo {title} {Computational fluid dynamics assessment of
  congenital tracheal stenosis},}\ }\href {\doibase 10.1007/s00383-022-05228-6}
  {\bibfield  {journal} {\bibinfo  {journal} {Pediatric Surgery International}\
  }\textbf {\bibinfo {volume} {38}} (\bibinfo {year} {2022}),\
  10.1007/s00383-022-05228-6}\BibitemShut {NoStop}%
\bibitem [{\citenamefont {R{\"u}ttgers}\ \emph {et~al.}(2021)\citenamefont
  {R{\"u}ttgers}, \citenamefont {Waldmann}, \citenamefont {Schr{\"o}der},\ and\
  \citenamefont {Lintermann}}]{Ruettgers2021}%
  \BibitemOpen
  \bibfield  {author} {\bibinfo {author} {\bibfnamefont {M.}~\bibnamefont
  {R{\"u}ttgers}}, \bibinfo {author} {\bibfnamefont {M.}~\bibnamefont
  {Waldmann}}, \bibinfo {author} {\bibfnamefont {W.}~\bibnamefont
  {Schr{\"o}der}}, \ and\ \bibinfo {author} {\bibfnamefont {A.}~\bibnamefont
  {Lintermann}},\ }\bibfield  {title} {\enquote {\bibinfo {title}
  {Machine-learning-based control of perturbed and heated channel flows},}\
  }in\ \href {\doibase 10.1007/978-3-030-90539-2_1} {\emph {\bibinfo
  {booktitle} {High Performance Computing}}},\ \bibinfo {editor} {edited by\
  \bibinfo {editor} {\bibfnamefont {H.}~\bibnamefont {Jagode}}, \bibinfo
  {editor} {\bibfnamefont {H.}~\bibnamefont {Anzt}}, \bibinfo {editor}
  {\bibfnamefont {H.}~\bibnamefont {Ltaief}}, \ and\ \bibinfo {editor}
  {\bibfnamefont {P.}~\bibnamefont {Luszczek}}}\ (\bibinfo  {publisher}
  {Springer International Publishing},\ \bibinfo {address} {Cham},\ \bibinfo
  {year} {2021})\ pp.\ \bibinfo {pages} {7--22}\BibitemShut {NoStop}%
\bibitem [{\citenamefont {Srivastava}\ and\ \citenamefont
  {Anand}(2024)}]{Vivek2024}%
  \BibitemOpen
  \bibfield  {author} {\bibinfo {author} {\bibfnamefont {V.}~\bibnamefont
  {Srivastava}}\ and\ \bibinfo {author} {\bibfnamefont {A.~R.}\ \bibnamefont
  {Anand}},\ }\enquote {\bibinfo {title} {Analysis of 2d human airway in
  laminar and turbulent flow model},}\ \ (\bibinfo {year} {2024})\ pp.\
  \bibinfo {pages} {855--863}\BibitemShut {NoStop}%
\bibitem [{\citenamefont {Johari}\ \emph
  {et~al.}(2013{\natexlab{a}})\citenamefont {Johari}, \citenamefont {Osman},
  \citenamefont {Helmi},\ and\ \citenamefont {Abdul~Kadir}}]{Johari2013}%
  \BibitemOpen
  \bibfield  {author} {\bibinfo {author} {\bibfnamefont {N.}~\bibnamefont
  {Johari}}, \bibinfo {author} {\bibfnamefont {K.}~\bibnamefont {Osman}},
  \bibinfo {author} {\bibfnamefont {N.}~\bibnamefont {Helmi}}, \ and\ \bibinfo
  {author} {\bibfnamefont {M.}~\bibnamefont {Abdul~Kadir}},\ }\bibfield
  {title} {\enquote {\bibinfo {title} {Comparative analysis of realistic
  ct-scan and simplified human airway models in airflow simulation},}\ }\href
  {\doibase 10.1080/10255842.2013.776548} {\bibfield  {journal} {\bibinfo
  {journal} {Computer methods in biomechanics and biomedical engineering}\
  }\textbf {\bibinfo {volume} {18}} (\bibinfo {year} {2013}{\natexlab{a}}),\
  10.1080/10255842.2013.776548}\BibitemShut {NoStop}%
\bibitem [{\citenamefont {{De Backer}}\ \emph {et~al.}(2008)\citenamefont {{De
  Backer}}, \citenamefont {Vos}, \citenamefont {Gorlé}, \citenamefont
  {Germonpré}, \citenamefont {Partoens}, \citenamefont {F.L.Wuyts},
  \citenamefont {Parizel},\ and\ \citenamefont {{De Backer}}}]{DeBacker2008}%
  \BibitemOpen
  \bibfield  {author} {\bibinfo {author} {\bibfnamefont {J.}~\bibnamefont {{De
  Backer}}}, \bibinfo {author} {\bibfnamefont {W.}~\bibnamefont {Vos}},
  \bibinfo {author} {\bibfnamefont {C.}~\bibnamefont {Gorlé}}, \bibinfo
  {author} {\bibfnamefont {P.}~\bibnamefont {Germonpré}}, \bibinfo {author}
  {\bibfnamefont {B.}~\bibnamefont {Partoens}}, \bibinfo {author} {\bibnamefont
  {F.L.Wuyts}}, \bibinfo {author} {\bibfnamefont {P.}~\bibnamefont {Parizel}},
  \ and\ \bibinfo {author} {\bibfnamefont {W.}~\bibnamefont {{De Backer}}},\
  }\bibfield  {title} {\enquote {\bibinfo {title} {Flow analyses in the lower
  airways: Patient-specific model and boundary conditions},}\ }\href {\doibase
  10.1016/j.medengphy.2007.11.002} {\bibfield  {journal} {\bibinfo  {journal}
  {Medical Engineering \& Physics}\ }\textbf {\bibinfo {volume} {30}},\
  \bibinfo {pages} {872--879} (\bibinfo {year} {2008})}\BibitemShut {NoStop}%
\bibitem [{\citenamefont {Aljawad}\ \emph {et~al.}(2021)\citenamefont
  {Aljawad}, \citenamefont {Rüttgers}, \citenamefont {Lintermann},
  \citenamefont {Schröder},\ and\ \citenamefont {Lee}}]{Aljawad2021}%
  \BibitemOpen
  \bibfield  {author} {\bibinfo {author} {\bibfnamefont {H.}~\bibnamefont
  {Aljawad}}, \bibinfo {author} {\bibfnamefont {M.}~\bibnamefont {Rüttgers}},
  \bibinfo {author} {\bibfnamefont {A.}~\bibnamefont {Lintermann}}, \bibinfo
  {author} {\bibfnamefont {W.}~\bibnamefont {Schröder}}, \ and\ \bibinfo
  {author} {\bibfnamefont {K.}~\bibnamefont {Lee}},\ }\bibfield  {title}
  {\enquote {\bibinfo {title} {Effects of the nasal cavity complexity on the
  pharyngeal airway fluid mechanics: A computational study},}\ }\href {\doibase
  10.1007/s10278-021-00501-x} {\bibfield  {journal} {\bibinfo  {journal}
  {Journal of Digital Imaging}\ }\textbf {\bibinfo {volume} {34}},\ \bibinfo
  {pages} {1120--1133} (\bibinfo {year} {2021})}\BibitemShut {NoStop}%
\bibitem [{\citenamefont {Johari}\ \emph
  {et~al.}(2013{\natexlab{b}})\citenamefont {Johari}, \citenamefont {Osman},
  \citenamefont {Helmi},\ and\ \citenamefont {Abdul~Kadir}}]{Nasrul2013}%
  \BibitemOpen
  \bibfield  {author} {\bibinfo {author} {\bibfnamefont {N.}~\bibnamefont
  {Johari}}, \bibinfo {author} {\bibfnamefont {K.}~\bibnamefont {Osman}},
  \bibinfo {author} {\bibfnamefont {N.}~\bibnamefont {Helmi}}, \ and\ \bibinfo
  {author} {\bibfnamefont {M.}~\bibnamefont {Abdul~Kadir}},\ }\bibfield
  {title} {\enquote {\bibinfo {title} {Comparative analysis of realistic
  ct-scan and simplified human airway models in airflow simulation},}\ }\href
  {\doibase 10.1080/10255842.2013.776548} {\bibfield  {journal} {\bibinfo
  {journal} {Computer methods in biomechanics and biomedical engineering}\
  }\textbf {\bibinfo {volume} {18}} (\bibinfo {year} {2013}{\natexlab{b}}),\
  10.1080/10255842.2013.776548}\BibitemShut {NoStop}%
\bibitem [{\citenamefont {Sommerfeld}\ \emph {et~al.}(2021)\citenamefont
  {Sommerfeld}, \citenamefont {Sgrott}, \citenamefont {Taborda}, \citenamefont
  {Koullapis}, \citenamefont {Bauer},\ and\ \citenamefont
  {Kassinos}}]{Sommerfeld2021}%
  \BibitemOpen
  \bibfield  {author} {\bibinfo {author} {\bibfnamefont {M.}~\bibnamefont
  {Sommerfeld}}, \bibinfo {author} {\bibfnamefont {O.}~\bibnamefont {Sgrott}},
  \bibinfo {author} {\bibfnamefont {M.}~\bibnamefont {Taborda}}, \bibinfo
  {author} {\bibfnamefont {P.}~\bibnamefont {Koullapis}}, \bibinfo {author}
  {\bibfnamefont {K.}~\bibnamefont {Bauer}}, \ and\ \bibinfo {author}
  {\bibfnamefont {S.}~\bibnamefont {Kassinos}},\ }\bibfield  {title} {\enquote
  {\bibinfo {title} {Analysis of flow field and turbulence predictions in a
  lung model applying rans and implications for particle deposition},}\ }\href
  {\doibase 10.1016/j.ejps.2021.105959} {\bibfield  {journal} {\bibinfo
  {journal} {European Journal of Pharmaceutical Sciences}\ }\textbf {\bibinfo
  {volume} {166}},\ \bibinfo {pages} {105959} (\bibinfo {year}
  {2021})}\BibitemShut {NoStop}%
\bibitem [{\citenamefont {Saksono}\ \emph {et~al.}(2011)\citenamefont
  {Saksono}, \citenamefont {Nithiarasu}, \citenamefont {Sazonov},\ and\
  \citenamefont {Yeo}}]{Saksono2011}%
  \BibitemOpen
  \bibfield  {author} {\bibinfo {author} {\bibfnamefont {P.~H.}\ \bibnamefont
  {Saksono}}, \bibinfo {author} {\bibfnamefont {P.}~\bibnamefont {Nithiarasu}},
  \bibinfo {author} {\bibfnamefont {I.}~\bibnamefont {Sazonov}}, \ and\
  \bibinfo {author} {\bibfnamefont {S.~Y.}\ \bibnamefont {Yeo}},\ }\bibfield
  {title} {\enquote {\bibinfo {title} {Computational flow studies in a
  subject-specific human upper airway using a one-equation turbulence model.
  influence of the nasal cavity},}\ }\href {\doibase 10.1002/nme.2986}
  {\bibfield  {journal} {\bibinfo  {journal} {International Journal for
  Numerical Methods in Engineering}\ }\textbf {\bibinfo {volume} {87}},\
  \bibinfo {pages} {96--114} (\bibinfo {year} {2011})}\BibitemShut {NoStop}%
\bibitem [{\citenamefont {Schillaci}\ and\ \citenamefont
  {Quadrio}(2022)}]{Schillaci2022}%
  \BibitemOpen
  \bibfield  {author} {\bibinfo {author} {\bibfnamefont {A.}~\bibnamefont
  {Schillaci}}\ and\ \bibinfo {author} {\bibfnamefont {M.}~\bibnamefont
  {Quadrio}},\ }\bibfield  {title} {\enquote {\bibinfo {title} {Importance of
  the numerical schemes in the cfd of the human nose},}\ }\href {\doibase
  10.1016/j.jbiomech.2022.111100} {\bibfield  {journal} {\bibinfo  {journal}
  {Journal of Biomechanics}\ }\textbf {\bibinfo {volume} {138}},\ \bibinfo
  {pages} {111100} (\bibinfo {year} {2022})}\BibitemShut {NoStop}%
\bibitem [{\citenamefont {Chen}\ and\ \citenamefont
  {Gutmark}(2013)}]{Chen2013}%
  \BibitemOpen
  \bibfield  {author} {\bibinfo {author} {\bibfnamefont {J.}~\bibnamefont
  {Chen}}\ and\ \bibinfo {author} {\bibfnamefont {E.}~\bibnamefont {Gutmark}},\
  }\bibfield  {title} {\enquote {\bibinfo {title} {Numerical investigation of
  airflow in an idealized human extra-thoracic airway: A comparison study},}\
  }\href {\doibase 10.1007/s10237-013-0496-x} {\bibfield  {journal} {\bibinfo
  {journal} {Biomechanics and modeling in mechanobiology}\ }\textbf {\bibinfo
  {volume} {13}} (\bibinfo {year} {2013}),\
  10.1007/s10237-013-0496-x}\BibitemShut {NoStop}%
\bibitem [{\citenamefont {Shao}\ \emph {et~al.}(2021)\citenamefont {Shao},
  \citenamefont {Yan}, \citenamefont {Liu},\ and\ \citenamefont
  {Lu}}]{Shao2021}%
  \BibitemOpen
  \bibfield  {author} {\bibinfo {author} {\bibfnamefont {J.}~\bibnamefont
  {Shao}}, \bibinfo {author} {\bibfnamefont {W.}~\bibnamefont {Yan}}, \bibinfo
  {author} {\bibfnamefont {Y.}~\bibnamefont {Liu}}, \ and\ \bibinfo {author}
  {\bibfnamefont {M.}~\bibnamefont {Lu}},\ }\bibfield  {title} {\enquote
  {\bibinfo {title} {Flow simulation in the upper respiratory tract of two
  obstructive sleep apnea patients with successful and failed surgery},}\
  }\href {\doibase 10.1155/2021/6683828} {\bibfield  {journal} {\bibinfo
  {journal} {Computational and Mathematical Methods in Medicine}\ }\textbf
  {\bibinfo {volume} {2021}},\ \bibinfo {pages} {1--12} (\bibinfo {year}
  {2021})}\BibitemShut {NoStop}%
\bibitem [{\citenamefont {Zuber}\ \emph {et~al.}(2012)\citenamefont {Zuber},
  \citenamefont {Abdullah}, \citenamefont {Ismail}, \citenamefont {Shuaib},
  \citenamefont {Sheikh Ab~Hamid},\ and\ \citenamefont {Ahmad}}]{Zuber2012}%
  \BibitemOpen
  \bibfield  {author} {\bibinfo {author} {\bibfnamefont {M.}~\bibnamefont
  {Zuber}}, \bibinfo {author} {\bibfnamefont {M.}~\bibnamefont {Abdullah}},
  \bibinfo {author} {\bibfnamefont {R.}~\bibnamefont {Ismail}}, \bibinfo
  {author} {\bibfnamefont {I.~L.}\ \bibnamefont {Shuaib}}, \bibinfo {author}
  {\bibfnamefont {S.}~\bibnamefont {Sheikh Ab~Hamid}}, \ and\ \bibinfo {author}
  {\bibfnamefont {K.}~\bibnamefont {Ahmad}},\ }\bibfield  {title} {\enquote
  {\bibinfo {title} {Review: A critical overview of limitations of cfd modeling
  in nasal airflow m zubair, mz abdullah, r ismail, il shuaib, sa hamid, ka
  ahmad journal of medical and biological engineering 32 (2), 3-10},}\ }\href
  {\doibase 10.5405/jmbe.948} {\bibfield  {journal} {\bibinfo  {journal}
  {Journal of Medical and Biological Engineering}\ }\textbf {\bibinfo {volume}
  {32}} (\bibinfo {year} {2012}),\ 10.5405/jmbe.948}\BibitemShut {NoStop}%
\bibitem [{\citenamefont {Li}\ \emph {et~al.}(2017)\citenamefont {Li},
  \citenamefont {Jiang}, \citenamefont {Dong},\ and\ \citenamefont
  {Zhao}}]{Li2017}%
  \BibitemOpen
  \bibfield  {author} {\bibinfo {author} {\bibfnamefont {C.}~\bibnamefont
  {Li}}, \bibinfo {author} {\bibfnamefont {J.}~\bibnamefont {Jiang}}, \bibinfo
  {author} {\bibfnamefont {H.}~\bibnamefont {Dong}}, \ and\ \bibinfo {author}
  {\bibfnamefont {K.}~\bibnamefont {Zhao}},\ }\bibfield  {title} {\enquote
  {\bibinfo {title} {Computational modeling and validation of human nasal
  airflow under various breathing conditions},}\ }\href {\doibase
  10.1016/j.jbiomech.2017.08.031} {\bibfield  {journal} {\bibinfo  {journal}
  {Journal of Biomechanics}\ }\textbf {\bibinfo {volume} {64}},\ \bibinfo
  {pages} {59--68} (\bibinfo {year} {2017})}\BibitemShut {NoStop}%
\bibitem [{\citenamefont {Calmet}\ \emph {et~al.}(2015)\citenamefont {Calmet},
  \citenamefont {Gambaruto}, \citenamefont {Bates}, \citenamefont {Vázquez},
  \citenamefont {Houzeaux},\ and\ \citenamefont {Doorly}}]{Calmet2015}%
  \BibitemOpen
  \bibfield  {author} {\bibinfo {author} {\bibfnamefont {H.}~\bibnamefont
  {Calmet}}, \bibinfo {author} {\bibfnamefont {A.}~\bibnamefont {Gambaruto}},
  \bibinfo {author} {\bibfnamefont {A.}~\bibnamefont {Bates}}, \bibinfo
  {author} {\bibfnamefont {M.}~\bibnamefont {Vázquez}}, \bibinfo {author}
  {\bibfnamefont {G.}~\bibnamefont {Houzeaux}}, \ and\ \bibinfo {author}
  {\bibfnamefont {D.}~\bibnamefont {Doorly}},\ }\bibfield  {title} {\enquote
  {\bibinfo {title} {Large-scale cfd simulations of the transitional and
  turbulent regime for the large human airways during rapid inhalation},}\
  }\href {\doibase 10.1016/j.compbiomed.2015.12.003} {\bibfield  {journal}
  {\bibinfo  {journal} {Computers in Biology and Medicine}\ }\textbf {\bibinfo
  {volume} {69}} (\bibinfo {year} {2015}),\
  10.1016/j.compbiomed.2015.12.003}\BibitemShut {NoStop}%
\bibitem [{\citenamefont {Farkas}\ \emph {et~al.}(2020)\citenamefont {Farkas},
  \citenamefont {Lizal}, \citenamefont {Jedelsky}, \citenamefont {Elcner},
  \citenamefont {Karas}, \citenamefont {Belka}, \citenamefont {Misik},\ and\
  \citenamefont {Jicha}}]{Farkas2020}%
  \BibitemOpen
  \bibfield  {author} {\bibinfo {author} {\bibfnamefont {{\'A}.}~\bibnamefont
  {Farkas}}, \bibinfo {author} {\bibfnamefont {F.}~\bibnamefont {Lizal}},
  \bibinfo {author} {\bibfnamefont {J.}~\bibnamefont {Jedelsky}}, \bibinfo
  {author} {\bibfnamefont {J.}~\bibnamefont {Elcner}}, \bibinfo {author}
  {\bibfnamefont {J.}~\bibnamefont {Karas}}, \bibinfo {author} {\bibfnamefont
  {M.}~\bibnamefont {Belka}}, \bibinfo {author} {\bibfnamefont
  {O.}~\bibnamefont {Misik}}, \ and\ \bibinfo {author} {\bibfnamefont
  {M.}~\bibnamefont {Jicha}},\ }\bibfield  {title} {\enquote {\bibinfo {title}
  {The role of the combined use of experimental and computational methods in
  revealing the differences between the micron-size particle deposition
  patterns in healthy and asthmatic subjects},}\ }\href {\doibase
  10.1016/j.jaerosci.2020.105582} {\bibfield  {journal} {\bibinfo  {journal}
  {Journal of Aerosol Science}\ }\textbf {\bibinfo {volume} {147}},\ \bibinfo
  {pages} {105582} (\bibinfo {year} {2020})}\BibitemShut {NoStop}%
\bibitem [{\citenamefont {Lizal}\ \emph {et~al.}(2020)\citenamefont {Lizal},
  \citenamefont {Elcner}, \citenamefont {Jedelsky}, \citenamefont {Maly},
  \citenamefont {Jicha}, \citenamefont {Farkas}, \citenamefont {Belka},
  \citenamefont {Rehak}, \citenamefont {Adam}, \citenamefont {Brinek},
  \citenamefont {Laznovsky}, \citenamefont {Zikmund},\ and\ \citenamefont
  {Kaiser}}]{Lizal2020}%
  \BibitemOpen
  \bibfield  {author} {\bibinfo {author} {\bibfnamefont {F.}~\bibnamefont
  {Lizal}}, \bibinfo {author} {\bibfnamefont {J.}~\bibnamefont {Elcner}},
  \bibinfo {author} {\bibfnamefont {J.}~\bibnamefont {Jedelsky}}, \bibinfo
  {author} {\bibfnamefont {M.}~\bibnamefont {Maly}}, \bibinfo {author}
  {\bibfnamefont {M.}~\bibnamefont {Jicha}}, \bibinfo {author} {\bibfnamefont
  {{\'A}.}~\bibnamefont {Farkas}}, \bibinfo {author} {\bibfnamefont
  {M.}~\bibnamefont {Belka}}, \bibinfo {author} {\bibfnamefont
  {Z.}~\bibnamefont {Rehak}}, \bibinfo {author} {\bibfnamefont
  {J.}~\bibnamefont {Adam}}, \bibinfo {author} {\bibfnamefont {A.}~\bibnamefont
  {Brinek}}, \bibinfo {author} {\bibfnamefont {J.}~\bibnamefont {Laznovsky}},
  \bibinfo {author} {\bibfnamefont {T.}~\bibnamefont {Zikmund}}, \ and\
  \bibinfo {author} {\bibfnamefont {J.}~\bibnamefont {Kaiser}},\ }\bibfield
  {title} {\enquote {\bibinfo {title} {The effect of oral and nasal breathing
  on the deposition of inhaled particles in upper and tracheobronchial
  airways},}\ }\href {\doibase 10.1016/j.jaerosci.2020.105649} {\bibfield
  {journal} {\bibinfo  {journal} {Journal of Aerosol Science}\ }\textbf
  {\bibinfo {volume} {150}},\ \bibinfo {pages} {105649} (\bibinfo {year}
  {2020})}\BibitemShut {NoStop}%
\bibitem [{\citenamefont {Sadafi}\ \emph {et~al.}(2024)\citenamefont {Sadafi},
  \citenamefont {Tousi}, \citenamefont {Backer},\ and\ \citenamefont
  {Backer}}]{Sadafi2024}%
  \BibitemOpen
  \bibfield  {author} {\bibinfo {author} {\bibfnamefont {H.}~\bibnamefont
  {Sadafi}}, \bibinfo {author} {\bibfnamefont {N.}~\bibnamefont {Tousi}},
  \bibinfo {author} {\bibfnamefont {W.}~\bibnamefont {Backer}}, \ and\ \bibinfo
  {author} {\bibfnamefont {J.}~\bibnamefont {Backer}},\ }\bibfield  {title}
  {\enquote {\bibinfo {title} {Validation of computational fluid dynamics
  models for airway deposition with spect data of the same population},}\
  }\href {\doibase 10.1038/s41598-024-56033-1} {\bibfield  {journal} {\bibinfo
  {journal} {Scientific Reports}\ }\textbf {\bibinfo {volume} {14}} (\bibinfo
  {year} {2024}),\ 10.1038/s41598-024-56033-1}\BibitemShut {NoStop}%
\bibitem [{\citenamefont {Hörschler}, \citenamefont {Schröder},\ and\
  \citenamefont {Meinke}(2010)}]{Hoerschler2010}%
  \BibitemOpen
  \bibfield  {author} {\bibinfo {author} {\bibfnamefont {I.}~\bibnamefont
  {Hörschler}}, \bibinfo {author} {\bibfnamefont {W.}~\bibnamefont
  {Schröder}}, \ and\ \bibinfo {author} {\bibfnamefont {M.}~\bibnamefont
  {Meinke}},\ }\bibfield  {title} {\enquote {\bibinfo {title} {On the
  assumption of steadiness of nasal cavity flow},}\ }\href {\doibase
  https://doi.org/10.1016/j.jbiomech.2009.12.008} {\bibfield  {journal}
  {\bibinfo  {journal} {Journal of Biomechanics}\ }\textbf {\bibinfo {volume}
  {43}},\ \bibinfo {pages} {1081--1085} (\bibinfo {year} {2010})}\BibitemShut
  {NoStop}%
\bibitem [{\citenamefont {Lintermann}, \citenamefont {Meinke},\ and\
  \citenamefont {Schr{\"o}der}(2012)}]{Lintermann2012}%
  \BibitemOpen
  \bibfield  {author} {\bibinfo {author} {\bibfnamefont {A.}~\bibnamefont
  {Lintermann}}, \bibinfo {author} {\bibfnamefont {M.}~\bibnamefont {Meinke}},
  \ and\ \bibinfo {author} {\bibfnamefont {W.}~\bibnamefont {Schr{\"o}der}},\
  }\bibfield  {title} {\enquote {\bibinfo {title} {Investigations of human
  nasal cavity flows based on a lattice-boltzmann method},}\ }in\ \href@noop {}
  {\emph {\bibinfo {booktitle} {High Performance Computing on Vector Systems
  2011}}},\ \bibinfo {editor} {edited by\ \bibinfo {editor} {\bibfnamefont
  {M.}~\bibnamefont {Resch}}, \bibinfo {editor} {\bibfnamefont
  {X.}~\bibnamefont {Wang}}, \bibinfo {editor} {\bibfnamefont {W.}~\bibnamefont
  {Bez}}, \bibinfo {editor} {\bibfnamefont {E.}~\bibnamefont {Focht}}, \bibinfo
  {editor} {\bibfnamefont {H.}~\bibnamefont {Kobayashi}}, \ and\ \bibinfo
  {editor} {\bibfnamefont {S.}~\bibnamefont {Roller}}}\ (\bibinfo  {publisher}
  {Springer Berlin Heidelberg},\ \bibinfo {address} {Berlin, Heidelberg},\
  \bibinfo {year} {2012})\ pp.\ \bibinfo {pages} {143--158}\BibitemShut
  {NoStop}%
\bibitem [{\citenamefont {Tsega}(2022)}]{Tsega2022}%
  \BibitemOpen
  \bibfield  {author} {\bibinfo {author} {\bibfnamefont {E.}~\bibnamefont
  {Tsega}},\ }\bibfield  {title} {\enquote {\bibinfo {title} {Cfd simulations
  of respiratory airflow in human upper airways response to walking and running
  for oral breathing condition},}\ }\href {\doibase
  10.1016/j.heliyon.2022.e10039} {\bibfield  {journal} {\bibinfo  {journal}
  {Heliyon}\ }\textbf {\bibinfo {volume} {8}},\ \bibinfo {pages} {e10039}
  (\bibinfo {year} {2022})}\BibitemShut {NoStop}%
\bibitem [{\citenamefont {Gaddam}\ and\ \citenamefont
  {Santhanakrishnan}(2021)}]{Gaddam2021}%
  \BibitemOpen
  \bibfield  {author} {\bibinfo {author} {\bibfnamefont {M.~G.}\ \bibnamefont
  {Gaddam}}\ and\ \bibinfo {author} {\bibfnamefont {A.}~\bibnamefont
  {Santhanakrishnan}},\ }\bibfield  {title} {\enquote {\bibinfo {title}
  {Effects of varying inhalation duration and respiratory rate on human airway
  flow},}\ }\href {\doibase 10.3390/fluids6060221} {\bibfield  {journal}
  {\bibinfo  {journal} {Fluids}\ }\textbf {\bibinfo {volume} {6}} (\bibinfo
  {year} {2021}),\ 10.3390/fluids6060221}\BibitemShut {NoStop}%
\bibitem [{m-A(2024)}]{m-AIA}%
  \BibitemOpen
  \href {\doibase 10.5281/zenodo.13350586} {\enquote {\bibinfo {title}
  {Institute of aerodynamics, m-aia},}\ } (\bibinfo {year} {2024})\BibitemShut
  {NoStop}%
\bibitem [{\citenamefont {Waldmann}\ \emph {et~al.}(2020)\citenamefont
  {Waldmann}, \citenamefont {Lintermann}, \citenamefont {Choi},\ and\
  \citenamefont {Schr{\"o}der}}]{Waldmann2020}%
  \BibitemOpen
  \bibfield  {author} {\bibinfo {author} {\bibfnamefont {M.}~\bibnamefont
  {Waldmann}}, \bibinfo {author} {\bibfnamefont {A.}~\bibnamefont
  {Lintermann}}, \bibinfo {author} {\bibfnamefont {Y.~J.}\ \bibnamefont
  {Choi}}, \ and\ \bibinfo {author} {\bibfnamefont {W.}~\bibnamefont
  {Schr{\"o}der}},\ }\bibfield  {title} {\enquote {\bibinfo {title} {Analysis
  of the effects of marme treatment on respiratory flow using the
  lattice-boltzmann method},}\ }in\ \href@noop {} {\emph {\bibinfo {booktitle}
  {New Results in Numerical and Experimental Fluid Mechanics XII}}},\ \bibinfo
  {editor} {edited by\ \bibinfo {editor} {\bibfnamefont {A.}~\bibnamefont
  {Dillmann}}, \bibinfo {editor} {\bibfnamefont {G.}~\bibnamefont {Heller}},
  \bibinfo {editor} {\bibfnamefont {E.}~\bibnamefont {Kr{\"a}mer}}, \bibinfo
  {editor} {\bibfnamefont {C.}~\bibnamefont {Wagner}}, \bibinfo {editor}
  {\bibfnamefont {C.}~\bibnamefont {Tropea}}, \ and\ \bibinfo {editor}
  {\bibfnamefont {S.}~\bibnamefont {Jakirli{\'{c}}}}}\ (\bibinfo  {publisher}
  {Springer International Publishing},\ \bibinfo {address} {Cham},\ \bibinfo
  {year} {2020})\ pp.\ \bibinfo {pages} {853--863}\BibitemShut {NoStop}%
\bibitem [{\citenamefont {Lintermann}\ and\ \citenamefont
  {Schröder}(2019)}]{Lintermann2019}%
  \BibitemOpen
  \bibfield  {author} {\bibinfo {author} {\bibfnamefont {A.}~\bibnamefont
  {Lintermann}}\ and\ \bibinfo {author} {\bibfnamefont {W.}~\bibnamefont
  {Schröder}},\ }\bibfield  {title} {\enquote {\bibinfo {title} {A
  hierarchical numerical journey through the nasal cavity: from nose-like
  models to real anatomies},}\ }\href {\doibase 10.1007/s10494-017-9876-0}
  {\bibfield  {journal} {\bibinfo  {journal} {Flow, Turbulence and Combustion}\
  }\textbf {\bibinfo {volume} {102}},\ \bibinfo {pages} {89--102} (\bibinfo
  {year} {2019})}\BibitemShut {NoStop}%
\bibitem [{\citenamefont {Waldmann}\ \emph {et~al.}(2021)\citenamefont
  {Waldmann}, \citenamefont {Grosch}, \citenamefont {Witzler}, \citenamefont
  {Lehner}, \citenamefont {Benda}, \citenamefont {Koch}, \citenamefont {Vogt},
  \citenamefont {Kohn}, \citenamefont {Schröder}, \citenamefont {Göbbert},\
  and\ \citenamefont {Lintermann}}]{Waldmann2021}%
  \BibitemOpen
  \bibfield  {author} {\bibinfo {author} {\bibfnamefont {M.}~\bibnamefont
  {Waldmann}}, \bibinfo {author} {\bibfnamefont {A.}~\bibnamefont {Grosch}},
  \bibinfo {author} {\bibfnamefont {C.}~\bibnamefont {Witzler}}, \bibinfo
  {author} {\bibfnamefont {M.}~\bibnamefont {Lehner}}, \bibinfo {author}
  {\bibfnamefont {O.}~\bibnamefont {Benda}}, \bibinfo {author} {\bibfnamefont
  {W.}~\bibnamefont {Koch}}, \bibinfo {author} {\bibfnamefont {K.}~\bibnamefont
  {Vogt}}, \bibinfo {author} {\bibfnamefont {C.}~\bibnamefont {Kohn}}, \bibinfo
  {author} {\bibfnamefont {W.}~\bibnamefont {Schröder}}, \bibinfo {author}
  {\bibfnamefont {J.}~\bibnamefont {Göbbert}}, \ and\ \bibinfo {author}
  {\bibfnamefont {A.}~\bibnamefont {Lintermann}},\ }\bibfield  {title}
  {\enquote {\bibinfo {title} {An effective simulation- and measurement-based
  workflow for enhanced diagnostics in rhinology},}\ }\href {\doibase
  10.1007/s11517-021-02446-3} {\bibfield  {journal} {\bibinfo  {journal}
  {Medical \& Biological Engineering \& Computing}\ }\textbf {\bibinfo {volume}
  {60}} (\bibinfo {year} {2021}),\ 10.1007/s11517-021-02446-3}\BibitemShut
  {NoStop}%
\bibitem [{\citenamefont {Rüttgers}\ \emph {et~al.}(2022)\citenamefont
  {Rüttgers}, \citenamefont {Waldmann}, \citenamefont {Schröder},\ and\
  \citenamefont {Lintermann}}]{Ruettgers2022}%
  \BibitemOpen
  \bibfield  {author} {\bibinfo {author} {\bibfnamefont {M.}~\bibnamefont
  {Rüttgers}}, \bibinfo {author} {\bibfnamefont {M.}~\bibnamefont {Waldmann}},
  \bibinfo {author} {\bibfnamefont {W.}~\bibnamefont {Schröder}}, \ and\
  \bibinfo {author} {\bibfnamefont {A.}~\bibnamefont {Lintermann}},\ }\bibfield
   {title} {\enquote {\bibinfo {title} {A machine-learning-based method for
  automatizing lattice-boltzmann simulations of respiratory flows},}\ }\href
  {\doibase 10.1007/s10489-021-02808-2} {\bibfield  {journal} {\bibinfo
  {journal} {Applied Intelligence}\ }\textbf {\bibinfo {volume} {52}},\
  \bibinfo {pages} {9080--9100} (\bibinfo {year} {2022})}\BibitemShut {NoStop}%
\bibitem [{\citenamefont {Liu}\ \emph {et~al.}(2024)\citenamefont {Liu},
  \citenamefont {Rüttgers}, \citenamefont {Quercia}, \citenamefont {Egele},
  \citenamefont {Pfaehler}, \citenamefont {Shende}, \citenamefont {Aach},
  \citenamefont {Schröder}, \citenamefont {Balaprakash},\ and\ \citenamefont
  {Lintermann}}]{Liu2024}%
  \BibitemOpen
  \bibfield  {author} {\bibinfo {author} {\bibfnamefont {X.}~\bibnamefont
  {Liu}}, \bibinfo {author} {\bibfnamefont {M.}~\bibnamefont {Rüttgers}},
  \bibinfo {author} {\bibfnamefont {A.}~\bibnamefont {Quercia}}, \bibinfo
  {author} {\bibfnamefont {R.}~\bibnamefont {Egele}}, \bibinfo {author}
  {\bibfnamefont {E.}~\bibnamefont {Pfaehler}}, \bibinfo {author}
  {\bibfnamefont {R.}~\bibnamefont {Shende}}, \bibinfo {author} {\bibfnamefont
  {M.}~\bibnamefont {Aach}}, \bibinfo {author} {\bibfnamefont {W.}~\bibnamefont
  {Schröder}}, \bibinfo {author} {\bibfnamefont {P.}~\bibnamefont
  {Balaprakash}}, \ and\ \bibinfo {author} {\bibfnamefont {A.}~\bibnamefont
  {Lintermann}},\ }\bibfield  {title} {\enquote {\bibinfo {title} {Refining
  computer tomography data with super-resolution networks to increase the
  accuracy of respiratory flow simulations},}\ }\href {\doibase
  10.1016/j.future.2024.05.020} {\bibfield  {journal} {\bibinfo  {journal}
  {Future Generation Computer Systems}\ }\textbf {\bibinfo {volume} {159}},\
  \bibinfo {pages} {474--488} (\bibinfo {year} {2024})}\BibitemShut {NoStop}%
\bibitem [{\citenamefont {Johanning-Meiners}\ \emph {et~al.}(2023)\citenamefont
  {Johanning-Meiners}, \citenamefont {Schier}, \citenamefont {Meyer},
  \citenamefont {D{\"o}rner}, \citenamefont {Gruhlke}, \citenamefont
  {Schr{\"o}der},\ and\ \citenamefont {Klaas}}]{JohanningMeiners.2023}%
  \BibitemOpen
  \bibfield  {author} {\bibinfo {author} {\bibfnamefont {B.~H.}\ \bibnamefont
  {Johanning-Meiners}}, \bibinfo {author} {\bibfnamefont {C.}~\bibnamefont
  {Schier}}, \bibinfo {author} {\bibfnamefont {T.}~\bibnamefont {Meyer}},
  \bibinfo {author} {\bibfnamefont {P.}~\bibnamefont {D{\"o}rner}}, \bibinfo
  {author} {\bibfnamefont {M.}~\bibnamefont {Gruhlke}}, \bibinfo {author}
  {\bibfnamefont {W.}~\bibnamefont {Schr{\"o}der}}, \ and\ \bibinfo {author}
  {\bibfnamefont {M.}~\bibnamefont {Klaas}},\ }\bibfield  {title} {\enquote
  {\bibinfo {title} {Combined velocity and aerosol deposition measurements in
  the respiratory tract using hs-piv and mtt assay},}\ }\href
  {https://scholarworks.calstate.edu/downloads/qb98mn684} {\bibfield  {journal}
  {\bibinfo  {journal} {15th International Symposium on Particle Image
  Velocimetry}\ } (\bibinfo {year} {2023})}\BibitemShut {NoStop}%
\bibitem [{\citenamefont {Johanning-Meiners}\ \emph {et~al.}(2024)\citenamefont
  {Johanning-Meiners}, \citenamefont {Mayolle}, \citenamefont {Schr{\"o}der},\
  and\ \citenamefont {Klaas}}]{JohanningMeiners.2024}%
  \BibitemOpen
  \bibfield  {author} {\bibinfo {author} {\bibfnamefont {B.~H.}\ \bibnamefont
  {Johanning-Meiners}}, \bibinfo {author} {\bibfnamefont {L.}~\bibnamefont
  {Mayolle}}, \bibinfo {author} {\bibfnamefont {W.}~\bibnamefont
  {Schr{\"o}der}}, \ and\ \bibinfo {author} {\bibfnamefont {M.}~\bibnamefont
  {Klaas}},\ }\bibfield  {title} {\enquote {\bibinfo {title} {3d lagrangian
  particle tracking in a model of the human airways},}\ }\href {\doibase
  10.55037/lxlaser.21st.22} {\bibfield  {journal} {\bibinfo  {journal}
  {Proceedings of the International Symposium on the Application of Laser and
  Imaging Techniques to Fluid Mechanics}\ }\textbf {\bibinfo {volume} {21}},\
  \bibinfo {pages} {1--13} (\bibinfo {year} {2024})}\BibitemShut {NoStop}%
\bibitem [{\citenamefont {Geller}\ \emph {et~al.}(2006)\citenamefont {Geller},
  \citenamefont {Krafczyk}, \citenamefont {Tölke}, \citenamefont {Turek},\
  and\ \citenamefont {Hron}}]{Geller2006}%
  \BibitemOpen
  \bibfield  {author} {\bibinfo {author} {\bibfnamefont {S.}~\bibnamefont
  {Geller}}, \bibinfo {author} {\bibfnamefont {M.}~\bibnamefont {Krafczyk}},
  \bibinfo {author} {\bibfnamefont {J.}~\bibnamefont {Tölke}}, \bibinfo
  {author} {\bibfnamefont {S.}~\bibnamefont {Turek}}, \ and\ \bibinfo {author}
  {\bibfnamefont {J.}~\bibnamefont {Hron}},\ }\bibfield  {title} {\enquote
  {\bibinfo {title} {Benchmark computations based on lattice-boltzmann, finite
  element and finite volume methods for laminar flows},}\ }\href {\doibase
  10.1016/j.compfluid.2005.08.009} {\bibfield  {journal} {\bibinfo  {journal}
  {Computers \& Fluids}\ }\textbf {\bibinfo {volume} {35}},\ \bibinfo {pages}
  {888--897} (\bibinfo {year} {2006})}\BibitemShut {NoStop}%
\bibitem [{\citenamefont {He}\ and\ \citenamefont
  {Luo}(1997)}]{heTheoryLatticeBoltzmann1997}%
  \BibitemOpen
  \bibfield  {author} {\bibinfo {author} {\bibfnamefont {X.}~\bibnamefont
  {He}}\ and\ \bibinfo {author} {\bibfnamefont {L.-S.}\ \bibnamefont {Luo}},\
  }\bibfield  {title} {\enquote {\bibinfo {title} {Theory of the lattice
  {{Boltzmann}} method: {{From}} the {{Boltzmann}} equation to the lattice
  {{Boltzmann}} equation},}\ }\href {\doibase 10.1103/PhysRevE.56.6811}
  {\bibfield  {journal} {\bibinfo  {journal} {Physical Review E}\ }\textbf
  {\bibinfo {volume} {56}},\ \bibinfo {pages} {6811--6817} (\bibinfo {year}
  {1997})}\BibitemShut {NoStop}%
\bibitem [{\citenamefont {Qian}, \citenamefont {D'Humi{\`e}res},\ and\
  \citenamefont {Lallemand}(1992)}]{qianLatticeBGKModels1992}%
  \BibitemOpen
  \bibfield  {author} {\bibinfo {author} {\bibfnamefont {Y.~H.}\ \bibnamefont
  {Qian}}, \bibinfo {author} {\bibfnamefont {D.}~\bibnamefont
  {D'Humi{\`e}res}}, \ and\ \bibinfo {author} {\bibfnamefont {P.}~\bibnamefont
  {Lallemand}},\ }\bibfield  {title} {\enquote {\bibinfo {title} {Lattice {{BGK
  Models}} for {{Navier-Stokes Equation}}},}\ }\href {\doibase
  10.1209/0295-5075/17/6/001} {\bibfield  {journal} {\bibinfo  {journal}
  {Europhysics Letters}\ }\textbf {\bibinfo {volume} {17}},\ \bibinfo {pages}
  {479--484} (\bibinfo {year} {1992})}\BibitemShut {NoStop}%
\bibitem [{\citenamefont {Bouzidi}, \citenamefont {Firdaouss},\ and\
  \citenamefont
  {Lallemand}(2001)}]{bouzidiMomentumTransferBoltzmannlattice2001}%
  \BibitemOpen
  \bibfield  {author} {\bibinfo {author} {\bibfnamefont {M.}~\bibnamefont
  {Bouzidi}}, \bibinfo {author} {\bibfnamefont {M.}~\bibnamefont {Firdaouss}},
  \ and\ \bibinfo {author} {\bibfnamefont {P.}~\bibnamefont {Lallemand}},\
  }\bibfield  {title} {\enquote {\bibinfo {title} {Momentum transfer of a
  {{Boltzmann-lattice}} fluid with boundaries},}\ }\href {\doibase
  10.1063/1.1399290} {\bibfield  {journal} {\bibinfo  {journal} {Physics of
  Fluids}\ }\textbf {\bibinfo {volume} {13}},\ \bibinfo {pages} {3452--3459}
  (\bibinfo {year} {2001})}\BibitemShut {NoStop}%
\bibitem [{\citenamefont {Lintermann}\ \emph {et~al.}(2014)\citenamefont
  {Lintermann}, \citenamefont {Schlimpert}, \citenamefont {Grimmen},
  \citenamefont {G{\"u}nther}, \citenamefont {Meinke},\ and\ \citenamefont
  {Schr{\"o}der}}]{Lintermann2014}%
  \BibitemOpen
  \bibfield  {author} {\bibinfo {author} {\bibfnamefont {A.}~\bibnamefont
  {Lintermann}}, \bibinfo {author} {\bibfnamefont {S.}~\bibnamefont
  {Schlimpert}}, \bibinfo {author} {\bibfnamefont {J.~H.}\ \bibnamefont
  {Grimmen}}, \bibinfo {author} {\bibfnamefont {C.}~\bibnamefont
  {G{\"u}nther}}, \bibinfo {author} {\bibfnamefont {M.}~\bibnamefont {Meinke}},
  \ and\ \bibinfo {author} {\bibfnamefont {W.}~\bibnamefont {Schr{\"o}der}},\
  }\bibfield  {title} {\enquote {\bibinfo {title} {Massively parallel grid
  generation on {{HPC}} systems},}\ }\href {\doibase 10.1016/j.cma.2014.04.009}
  {\bibfield  {journal} {\bibinfo  {journal} {Computer Methods in Applied
  Mechanics and Engineering}\ }\textbf {\bibinfo {volume} {277}},\ \bibinfo
  {pages} {131--153} (\bibinfo {year} {2014})}\BibitemShut {NoStop}%
\bibitem [{\citenamefont {Hartmann}, \citenamefont {Meinke},\ and\
  \citenamefont {Schröder}(2008)}]{Hartmann2008}%
  \BibitemOpen
  \bibfield  {author} {\bibinfo {author} {\bibfnamefont {D.}~\bibnamefont
  {Hartmann}}, \bibinfo {author} {\bibfnamefont {M.}~\bibnamefont {Meinke}}, \
  and\ \bibinfo {author} {\bibfnamefont {W.}~\bibnamefont {Schröder}},\
  }\bibfield  {title} {\enquote {\bibinfo {title} {An adaptive multilevel
  multigrid formulation for cartesian hierarchical grid methods},}\ }\href
  {\doibase 10.1016/j.compfluid.2007.06.007} {\bibfield  {journal} {\bibinfo
  {journal} {Computers \& Fluids}\ }\textbf {\bibinfo {volume} {37}},\ \bibinfo
  {pages} {1103--1125} (\bibinfo {year} {2008})}\BibitemShut {NoStop}%
\bibitem [{\citenamefont {Sagan}(1994)}]{Sagan1994}%
  \BibitemOpen
  \bibfield  {author} {\bibinfo {author} {\bibfnamefont {H.}~\bibnamefont
  {Sagan}},\ }\enquote {\bibinfo {title} {Hilbert's space-filling curve},}\ in\
  \href {\doibase 10.1007/978-1-4612-0871-6\_2} {\emph {\bibinfo {booktitle}
  {Space-Filling Curves}}}\ (\bibinfo  {publisher} {Springer New York},\
  \bibinfo {address} {New York, NY},\ \bibinfo {year} {1994})\ pp.\ \bibinfo
  {pages} {9--30}\BibitemShut {NoStop}%
\bibitem [{\citenamefont {Morton}(1966)}]{Morton1966}%
  \BibitemOpen
  \bibfield  {author} {\bibinfo {author} {\bibfnamefont {G.}~\bibnamefont
  {Morton}},\ }\href@noop {} {\emph {\bibinfo {title} {A Computer Oriented
  Geodetic Data Base and a New Technique in File Sequencing}}}\ (\bibinfo
  {publisher} {International Business Machines Company},\ \bibinfo {year}
  {1966})\BibitemShut {NoStop}%
\bibitem [{\citenamefont {Li}\ \emph {et~al.}(2003)\citenamefont {Li},
  \citenamefont {Liao}, \citenamefont {Choudhary}, \citenamefont {Ross},
  \citenamefont {Thakur}, \citenamefont {Gropp}, \citenamefont {Latham},
  \citenamefont {Siegel}, \citenamefont {Gallagher},\ and\ \citenamefont
  {Zingale}}]{Li2003}%
  \BibitemOpen
  \bibfield  {author} {\bibinfo {author} {\bibfnamefont {J.}~\bibnamefont
  {Li}}, \bibinfo {author} {\bibfnamefont {W.-K.}\ \bibnamefont {Liao}},
  \bibinfo {author} {\bibfnamefont {A.}~\bibnamefont {Choudhary}}, \bibinfo
  {author} {\bibfnamefont {R.}~\bibnamefont {Ross}}, \bibinfo {author}
  {\bibfnamefont {R.}~\bibnamefont {Thakur}}, \bibinfo {author} {\bibfnamefont
  {W.}~\bibnamefont {Gropp}}, \bibinfo {author} {\bibfnamefont
  {R.}~\bibnamefont {Latham}}, \bibinfo {author} {\bibfnamefont
  {A.}~\bibnamefont {Siegel}}, \bibinfo {author} {\bibfnamefont
  {B.}~\bibnamefont {Gallagher}}, \ and\ \bibinfo {author} {\bibfnamefont
  {M.}~\bibnamefont {Zingale}},\ }\bibfield  {title} {\enquote {\bibinfo
  {title} {Parallel netcdf: A high-performance scientific i/o interface},}\
  }in\ \href {\doibase 10.1109/SC.2003.10053} {\emph {\bibinfo {booktitle} {SC
  '03: Proceedings of the 2003 ACM/IEEE Conference on Supercomputing}}}\
  (\bibinfo {year} {2003})\ pp.\ \bibinfo {pages} {39--39}\BibitemShut
  {NoStop}%
\bibitem [{\citenamefont {Schanz}, \citenamefont {Gesemann},\ and\
  \citenamefont {Schr{\"o}der}(2016)}]{Schanz2016}%
  \BibitemOpen
  \bibfield  {author} {\bibinfo {author} {\bibfnamefont {D.}~\bibnamefont
  {Schanz}}, \bibinfo {author} {\bibfnamefont {S.}~\bibnamefont {Gesemann}}, \
  and\ \bibinfo {author} {\bibfnamefont {A.}~\bibnamefont {Schr{\"o}der}},\
  }\bibfield  {title} {\enquote {\bibinfo {title} {Shake-the-box: Lagrangian
  particle tracking at high particle image densities},}\ }\href {\doibase
  10.1007/s00348-016-2157-1} {\bibfield  {journal} {\bibinfo  {journal}
  {Experiments in Fluids}\ }\textbf {\bibinfo {volume} {57}} (\bibinfo {year}
  {2016}),\ 10.1007/s00348-016-2157-1}\BibitemShut {NoStop}%
\bibitem [{\citenamefont {Thörnig}(2021)}]{JURECA}%
  \BibitemOpen
  \bibfield  {author} {\bibinfo {author} {\bibfnamefont {P.}~\bibnamefont
  {Thörnig}},\ }\bibfield  {title} {\enquote {\bibinfo {title} {{JURECA: Data
  Centric and Booster Modules implementing the Modular Supercomputing
  Architecture at J\"{u}lich Supercomputing Centre}},}\ }\href {\doibase
  10.17815/jlsrf-7-182} {\bibfield  {journal} {\bibinfo  {journal} {Journal of
  large-scale research facilities}\ }\textbf {\bibinfo {volume} {7}} (\bibinfo
  {year} {2021}),\ 10.17815/jlsrf-7-182}\BibitemShut {NoStop}%
\bibitem [{\citenamefont {Wei}\ \emph {et~al.}(2024)\citenamefont {Wei},
  \citenamefont {He}, \citenamefont {Yang}, \citenamefont {Gu}, \citenamefont
  {Zhang}, \citenamefont {Sui}, \citenamefont {Zhou},\ and\ \citenamefont
  {Feng}}]{Wei2024}%
  \BibitemOpen
  \bibfield  {author} {\bibinfo {author} {\bibfnamefont {J.}~\bibnamefont
  {Wei}}, \bibinfo {author} {\bibfnamefont {X.}~\bibnamefont {He}}, \bibinfo
  {author} {\bibfnamefont {Q.}~\bibnamefont {Yang}}, \bibinfo {author}
  {\bibfnamefont {Q.}~\bibnamefont {Gu}}, \bibinfo {author} {\bibfnamefont
  {X.}~\bibnamefont {Zhang}}, \bibinfo {author} {\bibfnamefont
  {X.}~\bibnamefont {Sui}}, \bibinfo {author} {\bibfnamefont {R.}~\bibnamefont
  {Zhou}}, \ and\ \bibinfo {author} {\bibfnamefont {W.}~\bibnamefont {Feng}},\
  }\bibfield  {title} {\enquote {\bibinfo {title} {Numerical simulation of the
  influence of nasal cycle on nasal airflow},}\ }\href {\doibase
  10.1038/s41598-024-63024-9} {\bibfield  {journal} {\bibinfo  {journal}
  {Scientific Reports}\ }\textbf {\bibinfo {volume} {14}},\ \bibinfo {pages}
  {12161} (\bibinfo {year} {2024})}\BibitemShut {NoStop}%
\bibitem [{\citenamefont {Doty}(2001)}]{Doty2001}%
  \BibitemOpen
  \bibfield  {author} {\bibinfo {author} {\bibfnamefont {R.~L.}\ \bibnamefont
  {Doty}},\ }\bibfield  {title} {\enquote {\bibinfo {title} {Olfaction},}\
  }\href {\doibase 10.1146/annurev.psych.52.1.423} {\bibfield  {journal}
  {\bibinfo  {journal} {Annual Review of Psychology}\ }\textbf {\bibinfo
  {volume} {52}},\ \bibinfo {pages} {423--452} (\bibinfo {year}
  {2001})}\BibitemShut {NoStop}%
\bibitem [{\citenamefont {Dean}\ and\ \citenamefont {Hurst}(1959)}]{Dean1959}%
  \BibitemOpen
  \bibfield  {author} {\bibinfo {author} {\bibfnamefont {W.~R.}\ \bibnamefont
  {Dean}}\ and\ \bibinfo {author} {\bibfnamefont {J.~M.}\ \bibnamefont
  {Hurst}},\ }\bibfield  {title} {\enquote {\bibinfo {title} {Note on the
  motion of fluid in a curved pipe},}\ }\href {\doibase
  10.1112/S0025579300001947} {\bibfield  {journal} {\bibinfo  {journal}
  {Mathematika}\ }\textbf {\bibinfo {volume} {6}},\ \bibinfo {pages} {77--85}
  (\bibinfo {year} {1959})}\BibitemShut {NoStop}%
\bibitem [{\citenamefont {Y.Dubief}\ and\ \citenamefont
  {Delcayre}(2000)}]{Dubief00}%
  \BibitemOpen
  \bibfield  {author} {\bibinfo {author} {\bibnamefont {Y.Dubief}}\ and\
  \bibinfo {author} {\bibfnamefont {F.}~\bibnamefont {Delcayre}},\ }\bibfield
  {title} {\enquote {\bibinfo {title} {On coherent-vortex identification in
  turbulence},}\ }\href {\doibase 10.1088/1468-5248/1/1/011} {\bibfield
  {journal} {\bibinfo  {journal} {Journal of Turbulence}\ }\textbf {\bibinfo
  {volume} {1}},\ \bibinfo {pages} {N11} (\bibinfo {year} {2000})}\BibitemShut
  {NoStop}%
\bibitem [{\citenamefont {Krebs}\ \emph {et~al.}(2011)\citenamefont {Krebs},
  \citenamefont {Silva}, \citenamefont {Sciamarella},\ and\ \citenamefont
  {Artana}}]{Krebs2011}%
  \BibitemOpen
  \bibfield  {author} {\bibinfo {author} {\bibfnamefont {F.}~\bibnamefont
  {Krebs}}, \bibinfo {author} {\bibfnamefont {F.}~\bibnamefont {Silva}},
  \bibinfo {author} {\bibfnamefont {D.}~\bibnamefont {Sciamarella}}, \ and\
  \bibinfo {author} {\bibfnamefont {G.}~\bibnamefont {Artana}},\ }\bibfield
  {title} {\enquote {\bibinfo {title} {A three-dimensional study of the glottal
  jet},}\ }\href {\doibase 10.1007/s00348-011-1247-3} {\bibfield  {journal}
  {\bibinfo  {journal} {Experiments in Fluids}\ }\textbf {\bibinfo {volume}
  {52}} (\bibinfo {year} {2011}),\ 10.1007/s00348-011-1247-3}\BibitemShut
  {NoStop}%
\bibitem [{\citenamefont {Shinwari}\ \emph {et~al.}(2003)\citenamefont
  {Shinwari}, \citenamefont {Scherer}, \citenamefont {DeWitt},\ and\
  \citenamefont {Afjeh}}]{Shinwari2003}%
  \BibitemOpen
  \bibfield  {author} {\bibinfo {author} {\bibfnamefont {D.}~\bibnamefont
  {Shinwari}}, \bibinfo {author} {\bibfnamefont {R.~C.}\ \bibnamefont
  {Scherer}}, \bibinfo {author} {\bibfnamefont {K.~J.}\ \bibnamefont {DeWitt}},
  \ and\ \bibinfo {author} {\bibfnamefont {A.~A.}\ \bibnamefont {Afjeh}},\
  }\bibfield  {title} {\enquote {\bibinfo {title} {Flow visualization and
  pressure distributions in a model of the glottis with a symmetric and oblique
  divergent angle of 10 degrees},}\ }\href {\doibase 10.1121/1.1526468}
  {\bibfield  {journal} {\bibinfo  {journal} {The Journal of the Acoustical
  Society of America}\ }\textbf {\bibinfo {volume} {113}},\ \bibinfo {pages}
  {487--497} (\bibinfo {year} {2003})}\BibitemShut {NoStop}%
\bibitem [{\citenamefont {Zwicker}\ \emph {et~al.}(2017)\citenamefont
  {Zwicker}, \citenamefont {Ostilla-Mónico}, \citenamefont {Lieberman},\ and\
  \citenamefont {Brenner}}]{Zwicker2017}%
  \BibitemOpen
  \bibfield  {author} {\bibinfo {author} {\bibfnamefont {D.}~\bibnamefont
  {Zwicker}}, \bibinfo {author} {\bibfnamefont {R.}~\bibnamefont
  {Ostilla-Mónico}}, \bibinfo {author} {\bibfnamefont {D.}~\bibnamefont
  {Lieberman}}, \ and\ \bibinfo {author} {\bibfnamefont {M.}~\bibnamefont
  {Brenner}},\ }\bibfield  {title} {\enquote {\bibinfo {title} {Physical and
  geometric constraints explain the labyrinth-like shape of the nasal
  cavity},}\ }\href {\doibase 10.1073/pnas.1714795115} {\bibfield  {journal}
  {\bibinfo  {journal} {Proceedings of the National Academy of Sciences}\
  }\textbf {\bibinfo {volume} {115}} (\bibinfo {year} {2017}),\
  10.1073/pnas.1714795115}\BibitemShut {NoStop}%
\bibitem [{\citenamefont {Keck}\ \emph {et~al.}(2001)\citenamefont {Keck},
  \citenamefont {Leiacker}, \citenamefont {Heinrich}, \citenamefont
  {Kühnemann},\ and\ \citenamefont {Rettinger}}]{Keck2001}%
  \BibitemOpen
  \bibfield  {author} {\bibinfo {author} {\bibfnamefont {T.}~\bibnamefont
  {Keck}}, \bibinfo {author} {\bibfnamefont {R.}~\bibnamefont {Leiacker}},
  \bibinfo {author} {\bibfnamefont {A.}~\bibnamefont {Heinrich}}, \bibinfo
  {author} {\bibfnamefont {S.}~\bibnamefont {Kühnemann}}, \ and\ \bibinfo
  {author} {\bibfnamefont {G.}~\bibnamefont {Rettinger}},\ }\bibfield  {title}
  {\enquote {\bibinfo {title} {Humidity and temperature profile in the nasal
  cavity},}\ }\href@noop {} {\bibfield  {journal} {\bibinfo  {journal}
  {Rhinology}\ }\textbf {\bibinfo {volume} {38}},\ \bibinfo {pages} {167--71}
  (\bibinfo {year} {2001})}\BibitemShut {NoStop}%
\bibitem [{\citenamefont {Loring}, \citenamefont {Garcia-Jacques},\ and\
  \citenamefont {Malhotra}(2009)}]{Loring2009}%
  \BibitemOpen
  \bibfield  {author} {\bibinfo {author} {\bibfnamefont {S.}~\bibnamefont
  {Loring}}, \bibinfo {author} {\bibfnamefont {M.}~\bibnamefont
  {Garcia-Jacques}}, \ and\ \bibinfo {author} {\bibfnamefont {A.}~\bibnamefont
  {Malhotra}},\ }\bibfield  {title} {\enquote {\bibinfo {title} {Pulmonary
  characteristics in copd and mechanisms of increased work of breathing},}\
  }\href {\doibase 10.1152/japplphysiol.00008.2009} {\bibfield  {journal}
  {\bibinfo  {journal} {Journal of applied physiology (Bethesda, Md. : 1985)}\
  }\textbf {\bibinfo {volume} {107}},\ \bibinfo {pages} {309--14} (\bibinfo
  {year} {2009})}\BibitemShut {NoStop}%
\bibitem [{\citenamefont {Antiga}\ \emph {et~al.}(2008)\citenamefont {Antiga},
  \citenamefont {Piccinelli}, \citenamefont {Botti}, \citenamefont
  {Ene-Iordache}, \citenamefont {Remuzzi},\ and\ \citenamefont
  {Steinman}}]{Antiga2008}%
  \BibitemOpen
  \bibfield  {author} {\bibinfo {author} {\bibfnamefont {L.}~\bibnamefont
  {Antiga}}, \bibinfo {author} {\bibfnamefont {M.}~\bibnamefont {Piccinelli}},
  \bibinfo {author} {\bibfnamefont {L.}~\bibnamefont {Botti}}, \bibinfo
  {author} {\bibfnamefont {B.}~\bibnamefont {Ene-Iordache}}, \bibinfo {author}
  {\bibfnamefont {A.}~\bibnamefont {Remuzzi}}, \ and\ \bibinfo {author}
  {\bibfnamefont {D.~A.}\ \bibnamefont {Steinman}},\ }\bibfield  {title}
  {\enquote {\bibinfo {title} {An image-based modeling framework for
  patient-specific computational hemodynamics},}\ }\href {\doibase
  10.1007/s11517-008-0420-1} {\bibfield  {journal} {\bibinfo  {journal}
  {Medical \& Biological Engineering \& Computing}\ }\textbf {\bibinfo {volume}
  {46}},\ \bibinfo {pages} {1097--1112} (\bibinfo {year} {2008})}\BibitemShut
  {NoStop}%
\bibitem [{\citenamefont {Kelly}, \citenamefont {Prasad},\ and\ \citenamefont
  {Wexler}(2000)}]{Kelly2000}%
  \BibitemOpen
  \bibfield  {author} {\bibinfo {author} {\bibfnamefont {J.~T.}\ \bibnamefont
  {Kelly}}, \bibinfo {author} {\bibfnamefont {A.~K.}\ \bibnamefont {Prasad}}, \
  and\ \bibinfo {author} {\bibfnamefont {A.~S.}\ \bibnamefont {Wexler}},\
  }\bibfield  {title} {\enquote {\bibinfo {title} {Detailed flow patterns in
  the nasal cavity},}\ }\href {\doibase 10.1152/jappl.2000.89.1.323} {\bibfield
   {journal} {\bibinfo  {journal} {Journal of Applied Physiology}\ }\textbf
  {\bibinfo {volume} {89}},\ \bibinfo {pages} {323--337} (\bibinfo {year}
  {2000})}\BibitemShut {NoStop}%
\bibitem [{\citenamefont {Zhao}\ \emph {et~al.}(2006)\citenamefont {Zhao},
  \citenamefont {Dalton}, \citenamefont {Yang},\ and\ \citenamefont
  {Scherer}}]{Zhao2006}%
  \BibitemOpen
  \bibfield  {author} {\bibinfo {author} {\bibfnamefont {K.}~\bibnamefont
  {Zhao}}, \bibinfo {author} {\bibfnamefont {P.}~\bibnamefont {Dalton}},
  \bibinfo {author} {\bibfnamefont {G.}~\bibnamefont {Yang}}, \ and\ \bibinfo
  {author} {\bibfnamefont {P.}~\bibnamefont {Scherer}},\ }\bibfield  {title}
  {\enquote {\bibinfo {title} {Numerical modeling of turbulent and laminar
  airflow and odorant transport during sniffing in the human and rat nose},}\
  }\href {\doibase 10.1093/chemse/bjj008} {\bibfield  {journal} {\bibinfo
  {journal} {Chemical senses}\ }\textbf {\bibinfo {volume} {31}},\ \bibinfo
  {pages} {107--18} (\bibinfo {year} {2006})}\BibitemShut {NoStop}%
\bibitem [{\citenamefont {Asgharian}, \citenamefont {Hofmann},\ and\
  \citenamefont {Bergmann}(2001)}]{Asgharian2001}%
  \BibitemOpen
  \bibfield  {author} {\bibinfo {author} {\bibfnamefont {B.}~\bibnamefont
  {Asgharian}}, \bibinfo {author} {\bibfnamefont {W.}~\bibnamefont {Hofmann}},
  \ and\ \bibinfo {author} {\bibfnamefont {R.}~\bibnamefont {Bergmann}},\
  }\bibfield  {title} {\enquote {\bibinfo {title} {Particle deposition in a
  multiple-path model of the human lung},}\ }\href {\doibase
  10.1080/02786820151092478} {\bibfield  {journal} {\bibinfo  {journal}
  {Aerosol Science and Technology - AEROSOL SCI TECH}\ }\textbf {\bibinfo
  {volume} {34}},\ \bibinfo {pages} {332--339} (\bibinfo {year}
  {2001})}\BibitemShut {NoStop}%
\bibitem [{\citenamefont {Zhang}, \citenamefont {Kleinstreuer},\ and\
  \citenamefont {Kim}(2001)}]{Zhang2001}%
  \BibitemOpen
  \bibfield  {author} {\bibinfo {author} {\bibfnamefont {Z.}~\bibnamefont
  {Zhang}}, \bibinfo {author} {\bibfnamefont {C.}~\bibnamefont {Kleinstreuer}},
  \ and\ \bibinfo {author} {\bibfnamefont {C.~S.}\ \bibnamefont {Kim}},\
  }\bibfield  {title} {\enquote {\bibinfo {title} {Effects of curved inlet
  tubes on air flow and particle deposition in bifurcating lung models},}\
  }\href {\doibase 10.1016/S0021-9290(00)00233-5} {\bibfield  {journal}
  {\bibinfo  {journal} {Journal of biomechanics}\ }\textbf {\bibinfo {volume}
  {34}},\ \bibinfo {pages} {659--69} (\bibinfo {year} {2001})}\BibitemShut
  {NoStop}%
\bibitem [{\citenamefont {Kolmogorov}\ \emph {et~al.}(1991)\citenamefont
  {Kolmogorov}, \citenamefont {Levin}, \citenamefont {Hunt}, \citenamefont
  {Phillips},\ and\ \citenamefont {Williams}}]{Kolmogorov1991}%
  \BibitemOpen
  \bibfield  {author} {\bibinfo {author} {\bibfnamefont {A.~N.}\ \bibnamefont
  {Kolmogorov}}, \bibinfo {author} {\bibfnamefont {V.}~\bibnamefont {Levin}},
  \bibinfo {author} {\bibfnamefont {J.~C.~R.}\ \bibnamefont {Hunt}}, \bibinfo
  {author} {\bibfnamefont {O.~M.}\ \bibnamefont {Phillips}}, \ and\ \bibinfo
  {author} {\bibfnamefont {D.}~\bibnamefont {Williams}},\ }\bibfield  {title}
  {\enquote {\bibinfo {title} {The local structure of turbulence in
  incompressible viscous fluid for very large reynolds numbers},}\ }\href
  {\doibase 10.1098/rspa.1991.0075} {\bibfield  {journal} {\bibinfo  {journal}
  {Proceedings of the Royal Society of London. Series A: Mathematical and
  Physical Sciences}\ }\textbf {\bibinfo {volume} {434}},\ \bibinfo {pages}
  {9--13} (\bibinfo {year} {1991})}\BibitemShut {NoStop}%
\bibitem [{\citenamefont {Schroter}\ and\ \citenamefont
  {Sudlow}(1969)}]{Schroter1969}%
  \BibitemOpen
  \bibfield  {author} {\bibinfo {author} {\bibfnamefont {R.}~\bibnamefont
  {Schroter}}\ and\ \bibinfo {author} {\bibfnamefont {M.}~\bibnamefont
  {Sudlow}},\ }\bibfield  {title} {\enquote {\bibinfo {title} {Flow patterns in
  models of the human bronchial airways},}\ }\href {\doibase
  10.1016/0034-5687(69)90018-8} {\bibfield  {journal} {\bibinfo  {journal}
  {Respiration Physiology}\ }\textbf {\bibinfo {volume} {7}},\ \bibinfo {pages}
  {341--355} (\bibinfo {year} {1969})}\BibitemShut {NoStop}%
\bibitem [{\citenamefont {Soni}, \citenamefont {Lindley},\ and\ \citenamefont
  {Thompson}(2009)}]{Soni2009}%
  \BibitemOpen
  \bibfield  {author} {\bibinfo {author} {\bibfnamefont {B.}~\bibnamefont
  {Soni}}, \bibinfo {author} {\bibfnamefont {C.}~\bibnamefont {Lindley}}, \
  and\ \bibinfo {author} {\bibfnamefont {D.}~\bibnamefont {Thompson}},\
  }\bibfield  {title} {\enquote {\bibinfo {title} {The combined effects of
  non-planarity and asymmetry on primary and secondary flows in the small
  bronchial tubes},}\ }\href {\doibase 10.1002/fld.1802} {\bibfield  {journal}
  {\bibinfo  {journal} {International Journal for Numerical Methods in Fluids}\
  }\textbf {\bibinfo {volume} {59}},\ \bibinfo {pages} {117--146} (\bibinfo
  {year} {2009})}\BibitemShut {NoStop}%
\bibitem [{\citenamefont {Niemöller}\ \emph {et~al.}(2020)\citenamefont
  {Niemöller}, \citenamefont {Schlottke-Lakemper}, \citenamefont {Meinke},\
  and\ \citenamefont {Schröder}}]{Niemoeller2020}%
  \BibitemOpen
  \bibfield  {author} {\bibinfo {author} {\bibfnamefont {A.}~\bibnamefont
  {Niemöller}}, \bibinfo {author} {\bibfnamefont {M.}~\bibnamefont
  {Schlottke-Lakemper}}, \bibinfo {author} {\bibfnamefont {M.}~\bibnamefont
  {Meinke}}, \ and\ \bibinfo {author} {\bibfnamefont {W.}~\bibnamefont
  {Schröder}},\ }\bibfield  {title} {\enquote {\bibinfo {title} {Dynamic load
  balancing for direct-coupled multiphysics simulations},}\ }\href {\doibase
  10.1016/j.compfluid.2020.104437} {\bibfield  {journal} {\bibinfo  {journal}
  {Computers \& Fluids}\ }\textbf {\bibinfo {volume} {199}},\ \bibinfo {pages}
  {104437} (\bibinfo {year} {2020})}\BibitemShut {NoStop}%
\bibitem [{\citenamefont {Pendolino}\ \emph {et~al.}(2018)\citenamefont
  {Pendolino}, \citenamefont {Lund}, \citenamefont {Nardello},\ and\
  \citenamefont {Ottaviano}}]{Pendolino2018}%
  \BibitemOpen
  \bibfield  {author} {\bibinfo {author} {\bibfnamefont {A.}~\bibnamefont
  {Pendolino}}, \bibinfo {author} {\bibfnamefont {V.}~\bibnamefont {Lund}},
  \bibinfo {author} {\bibfnamefont {E.}~\bibnamefont {Nardello}}, \ and\
  \bibinfo {author} {\bibfnamefont {G.}~\bibnamefont {Ottaviano}},\ }\bibfield
  {title} {\enquote {\bibinfo {title} {The nasal cycle: a comprehensive
  review*},}\ }\href {\doibase 10.4193/RHINOL/18.021} {\bibfield  {journal}
  {\bibinfo  {journal} {Rhinology}\ }\textbf {\bibinfo {volume} {1}} (\bibinfo
  {year} {2018}),\ 10.4193/RHINOL/18.021}\BibitemShut {NoStop}%
\end{thebibliography}%

\end{document}